# Transparent framework to assess the revision of climate pledges after the first Global Stocktake

Kushal Tibrewal[1]†, Katsumasa Tanaka[1,2]*†, Philippe Ciais[1], Olivier Boucher[3]

[1]Laboratoire des Sciences du Climat et de l'Environnement (LSCE), IPSL, CEA/CNRS/UVSQ, Université Paris-Saclay; Gif-sur-Yvette, France

[2]Earth System Division, National Institute for Environmental Studies (NIES); Tsukuba, Japan

[3]Institut Pierre-Simon Laplace (IPSL), Sorbonne Université / CNRS; Paris, France

*Corresponding author. Email: katsumasa.tanaka@lsce.ipsl.fr

†These authors contributed equally to this work.

**Abstract:** To assess the impact of potential future climate pledges after the first Global Stocktake, we propose a simple, transparent framework for developing emission and temperature scenarios by country. We show that current pledges with unconditional targets lead to global warming of 1.96 (1.39-2.6)°C by 2100. Further warming could be limited through i) commitment to mid-century net-zero targets for all countries and earlier net-zero targets for developed countries, ii) enhancement of the Global Methane Pledge, and iii) ambitious implementation of the Glasgow Leaders' Declaration on Forests and Land Use. Our analysis further shows that overshooting 1.5°C is unavoidable, even with supplementary climate engineering strategies, suggesting the need for strategies to limit further overshoot and ultimately reduce the warming towards 1.5°C.

**One-Sentence Summary:** Overshooting 1.5°C is unavoidable, requiring strategies to limit further overshoot and reduce the warming towards 1.5°C.

**Main text**

The Paris Agreement stands as a cornerstone of global cooperation to address the need for mitigating climate change. Central to the Agreement are the Nationally Determined Contributions (NDCs) and Long-Term Strategies (LTSs) that individual countries have committed to, reflecting the collective ambition to hold global warming to well below 2°C above pre-industrial levels and pursue efforts to limit warming to 1.5°C. As the world confronts the urgency of climate action, there is a growing body of research scrutinizing these commitments.

Most recent studies have focused on modeling future emissions and temperatures to assess the effectiveness of these pledges prior to the 28th Conference of the Parties (COP28), where the first Global Stocktake (GST) was concluded (*1–4*). Several earlier studies were conducted to evaluate the pledges up to around COP26 (*5–9*). Studies generally indicate that if all NDCs and mid-century net-zero targets are realized, global warming can be held around 2°C by the end of the century. In contrast, with only pledges that are credible for implementation, the end-of-century warming can exceed 2°C by a wide margin. Following the outcome of GST at COP28, countries have approximately one year to further revise their current NDCs and another year to submit new 2035 NDCs (Paragraphs 166 and 170 of (*10*), respectively).

To explore strategies for enhancing national climate pledges, we consider NDCs and LTSs available by the end of COP28, as well as the Global Methane Pledge (GMP), which aims to reduce global anthropogenic $CH_4$ emissions by up to 30% by 2030 (*11*). Our analysis evaluates the potential impact of more ambitious targets, such as strengthened near-term mitigation, accelerated timelines for net-zero, expanded scope of GMP, and implementations of the Glasgow Leaders' Declaration on Forests and Land Use (LUF) (*12*). Reflecting on the discussions and outcome of GST, as well as the preceding Sunnylands Statement between US and China (*13*), we also assess potential non-$CO_2$ targets that appeared in the COP28 draft negotiation texts as they may be reconsidered at future COPs: i) net-zero for developed countries by 2040, ii) 40% $CH_4$ reduction by 2035, and iii) 13% $N_2O$ reduction by 2030, followed by 18% by 2035 (Paragraph 39 of (*14*)). We further investigate the adverse consequences of falling short of declared targets, as implementation of current pledges is key but sometimes doubtful. Our study goes beyond existing studies in that it isolates the temperature response to each upgrade and downgrade at the country level. Sectoral targets are not considered in this study.

Here we propose a simple framework for developing national emission scenarios for all Paris signatories in order to bring greater transparency to the ongoing discourse on the adequacy of current efforts to pursue the Paris objectives. There are large uncertainties in emission pathways after mid-century, influencing end-of-century temperatures, among other sources of uncertainty. Most previous studies either assumed a decline to net negative emissions after net-zero (for $CO_2$ or greenhouse gases (GHGs)) or were not very explicit about emission assumptions beyond net-zero. To our knowledge, continuing emission reductions after net-zero, which are not stated in NDCs, are directly or indirectly informed by the IPCC Sixth Assessment Report (AR6) scenario database, which contains a number of deep negative $CO_2$ emission scenarios, for which the Integrated Assessment Model (IAM) approach has been criticized (*15*). Building on Benveniste study (*16*), we use a simple mathematical function (*17*) to extrapolate emissions to zero $CO_2$ or GHG emissions for countries that have already committed to net-zero (Methods). Emissions of a country with only a 2030 target are assumed to decline no further after 2030 (i.e., Constant Emissions approach (*18*)). Our analysis focuses on three main gases ($CO_2$, $CH_4$, and $N_2O$) as GHGs. The IAM community produced emission pathways for selected countries using more detailed approaches (*19*) and called for the development of a standardized methodological framework for national scenarios (*20*).

Our temperature calculation method combines three well-established concepts: i) Global Warming Potential-star (GWP*) (*21–25*) and 100-year GWP (GWP100) (*26–29*) to convert the emissions of short-lived (e.g., $CH_4$) and long-lived (e.g., $N_2O$) climate forcers (*30*), respectively, to $CO_2$eq emissions, ii) Transient climate response to cumulative carbon emissions (TCRE) (*31–34*) to relate cumulative $CO_2$eq emissions to temperature changes, iii) Zero emissions commitment (ZEC) (*34–37*) to incorporate long-term temperature adjustments after net-zero (Methods). Our method is further supplemented and supported by a reduced-complexity climate model ACC2 (*38–40*) and the AR5 impulse response function (IRF) (*41*). Our emission and temperature pathways are publicly available and kept fully traceable.

While the primary focus remains on reducing emissions from existing sources, there is a growing recognition of the potential importance of carbon dioxide removal (CDR) and solar radiation modification (SRM) techniques as supplementary strategies (*42–45*). Our study considers three "novel" CDRs (*46*): bioenergy with carbon capture and storage (BECCS), direct air carbon capture and storage (DACCS) and enhanced weathering (*47*). We analyze the potential role of novel CDR and SRM in conjunction with climate pledges (*48*), in the context of concerns about the scalability of CDR and controversies on SRM due to their potential unintended consequences, termination issues and governance and ethical considerations (*49*, *50*).

**Projections under current national climate pledges**

As of December 2023, 194 countries have near-term targets (*51*), 93 have long-term targets to become carbon or GHG neutral (*52*), 150 have committed to GMP and over 140 have pledged to LUF. With all these commitments combined, global emissions are projected to peak at 51-53 $GtCO_2$eq yr$^{-1}$ around 2026, followed by a decline and a stabilization at 9-12 $GtCO_2$eq yr$^{-1}$ by 2070 (Fig. 1a). Emissions from the energy, industrial processes, waste and agricultural sectors (excluding the land use, land-use change, and forestry (LULUCF) sector) are reduced by 72% in 2060 compared to 2022. $CO_2$ continues to dominate GHG emissions, with an increased share of $CH_4$ from mid-century.

We analyzed the emission trajectories for individual countries/regions (Fig. S3) to estimate their relative contributions to emission budgets during recent historical (1990-2023), mid-century (2023-2060) and end-of-century (2061-2100) periods. Major developed countries (USA, EU27, Japan, Canada and UK) contributed 33% to the historical budget. As a result of the current pledges, their contributions are reduced to one-third in the mid-century and nearly zero in the end-of-century budget. China dominates the global $CO_2$eq budget, amounting to 22% and 34% for the historical and mid-century budgets, respectively. China's share drops to 18% in the end-of-century budget owing to its net-zero commitment for 2060. The emission budget during the mid- and end of the century is also strongly influenced by a group of developing/transitioning countries including EMME, India, Pakistan, Russia, Indonesia, Vietnam and Mexico, representing a combined share of 33% and 41% in the mid- and end-of-century budget, respectively, compared to 22% in the historical budget. Within this group, EMME accounts for the largest share of the end-of-century budget (18%), comparable to the share of China.

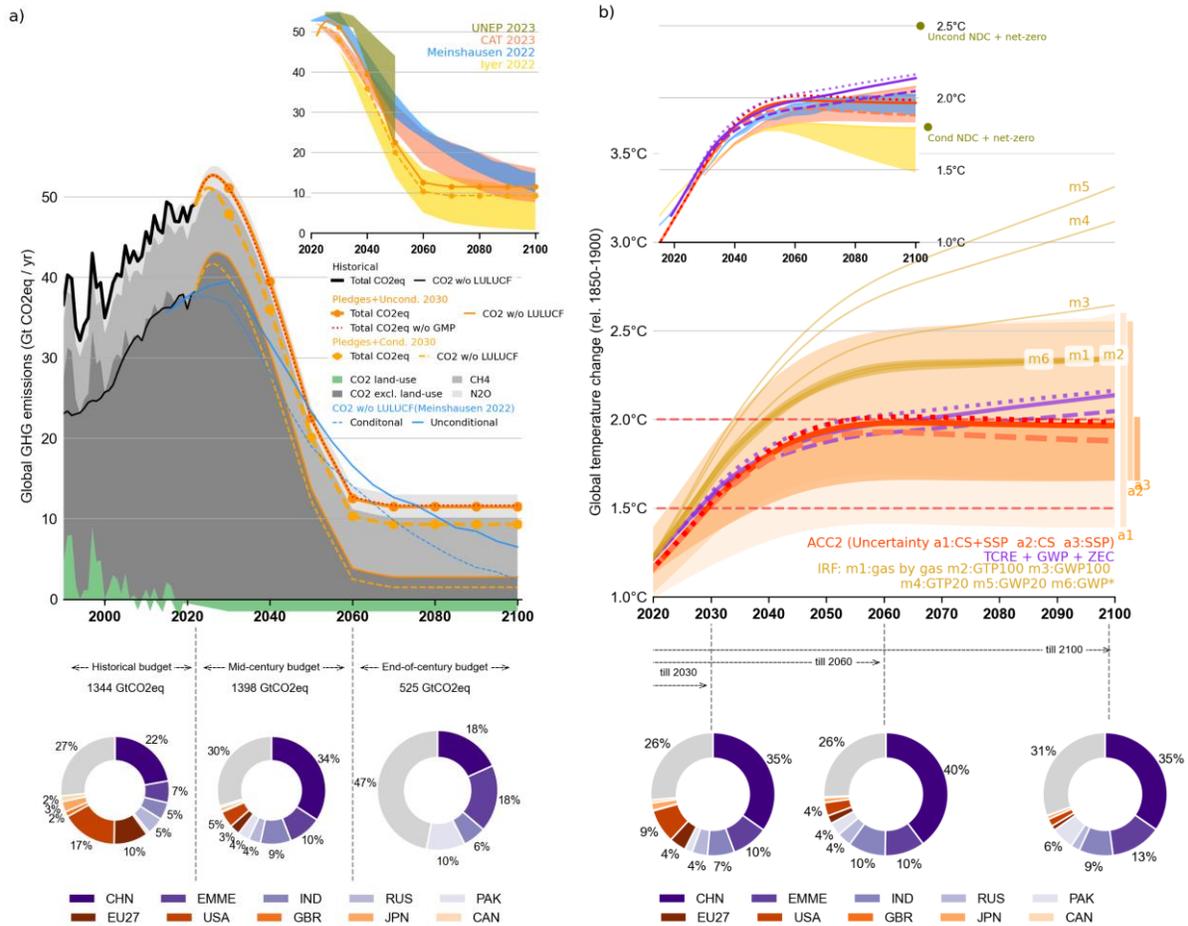

**Fig. 1. Emissions and temperature projection under current pledges (near-term targets, net-zeros and GMP). (a)** shows the global $CO_2$, $CH_4$ and $N_2O$ emissions from 1990 to 2100 (including LULUCF; GHG aggregation based on GWP100 (AR5 values following the Paris Agreement rulebook (*29*))). Lines represent the total annual $CO_2eq$ emissions from all sectors. Shaded areas in grey represent the contributions from individual gases (excluding LULUCF). Green represents LULUCF (net $CO_2$ fluxes over managed land), which include emissions from deforestation & degradation and sinks induced from environmental changes. Pie charts show the contributions of selected developed and developing countries/regions to cumulative $CO_2eq$ budgets (excluding LULUCF) for three periods. Regions are 27 countries of the European Union (EU27) and 16 countries of the Eastern Mediterranean and Middle East (EMME) area (following the definition of the EMME Climate Change Initiative): Bahrain, Egypt, Iran, Iraq, Israel, Jordan, Kuwait, Lebanon, Oman, Palestine, Qatar, Saudi Arabia, Syria, Turkey, the United Arab Emirates (UAE) and Yemen (Cyprus and Greece are considered only as part of EU27 in our analysis). **(b)** shows the global temperature trajectories calculated from ACC2 (red), IRF+metric (orange) and TCRE+GWP+ZEC (purple) approaches. Solid and dashed lines represent temperature projections under current pledges with unconditional and conditional targets, respectively (not shown for IRF+metric). Dotted lines indicate those without GMP, which are by default included in the current pledges in our analysis. Bars denote uncertainties calculated from ACC2, where "a1" includes the combined effect of the uncertainty in climate sensitivity from 2.5 to 4°C and the range of SSPs used for gases and pollutants other than $CO_2$, $CH_4$ and $N_2O$ (SSP1-1.9 by default), "a2" includes only the effect from climate sensitivity and "a3" includes only the effect from SSPs. Pie charts show the contributions for selected countries/regions to the temperature in 2030, 2060 and 2100 relative to 2019 based on country level emissions (excluding LULUCF) and TCRE+GWP+ZEC approach. Inserts in (a) and (b) compare our emission and temperature trajectories with those from other studies (*1*, *4*, *8*, *9*).

Using the TCRE+GWP+ZEC method, we study the temperature implications of the aforementioned emission pathways (Fig. 1b). The temperature rises rapidly up until 2030 due to i) strong growth of $CO_2$ emissions without LULUCF (derived from our approach) and ii) reduction in pollutant, particularly $SO_2$, emissions in SSP1-1.9 (53) used for gases and pollutants other than $CO_2$, $CH_4$, and $N_2O$ (unrelated to national climate pledges). Global warming is projected to reach 2.14°C and 2.05°C by the end of the century when considering unconditional and conditional targets, respectively. These temperature responses are comparable to those using ACC2 (38–40) (1.96 (1.39-2.6)°C and 1.88 (1.31-2.51)°C, respectively, with uncertainty ranges in parentheses), but they are lower than those that using IRF with GWP*. Our end-of-century projections are comparable to other studies, although certain underlying assumptions may vary (Methods).

China contributes the largest to the temperatures, followed by EMME and India across all time points. China causes an increase of 0.25°C by 2100 relative to 2019 levels with a crucial amount of warming during 2020-2040. This estimate is larger than IAM estimates for the 1.5°C target (54) also because of deep negative $CO_2$ emissions beyond net-zero typically assumed (55). EMME contributes a combined share of 10% to warming up to 2030 and 2060 each, which increases to 13% in 2100, or warming of 0.1°C. Since a substantial fraction of these countries do not have any long-term strategy, their contribution amplifies during the latter half of the century. India contributes 7-10% of warming at all time points, or warming of 0.06°C by 2100.

**Impacts of enhanced or delayed national climate pledges**

We further explored a suite of emissions pathways with enhanced pledges by country (Table 1 and Figs. 2, S4, and S5) to identify key potential enhancements for next NDC revisions. Inspired by the seminal work of Pacala and Socolow (56), we present such pledges as wedges. Note, however, that our wedges are not always mutually exclusive. In other words, the actual benefit of a wedge may depend to some extent on the presence of another wedge (the overlaps of wedges).

Ratcheting up the near-term conditional reduction targets by 10% (NDC01) (e.g., a target of X% increases to 1.1*X%) could lead to further reductions of 5 $GtCO_2eq$ $yr^{-1}$ in 2030 and 2050 each, saving 220 $GtCO_2eq$ during 2023-2100 and avoiding warming of 0.1°C. China, India and USA can provide the maximum benefit by increasing their near-term targets. On the contrary, failing to achieve their unconditional targets by 10% (NDC02) could increase emissions by 5 $GtCO_2eq/yr$ in 2030 and 2050 each, adding 200 $GtCO_2eq$ during 2023-2100 with warming of 0.1°C.

EU27 and 52 other countries proposed to achieve neutrality by 2050 and 14 aim to achieve it no later than 2070. If all these countries strive to achieve it by 2050 (LTS02), this would reduce 11 $GtCO_2eq/yr$ in 2050 and save 180 $GtCO_2eq$ during 2023-2070, avoiding warming of 0.08°C. China, India and Indonesia can provide the maximum benefit here. If countries with a net-zero target by 2050 (mainly developed countries) raise their ambition to achieve it by 2040 (LTS03) as in the COP28 draft negotiation text (14), it would reduce 7 $GtCO_2eq/yr$ in 2040 and 65 $GtCO_2eq$ during 2023-2050, avoiding 0.03°C. Furthermore, among countries with a net-zero target, eight countries have proposed a $CO_2$ only target and eight have an unclear target (assumed $CO_2$ only). If these also shift to a GHG net-zero (LTS05), it would reduce 3 $GtCO_2eq/yr$ in 2050 and 203 $GtCO_2eq$ during 2023-2100, avoiding 0.09°C. On the contrary, if all countries delay their respective neutrality target by 10 years (LTS04), it would increase emissions by 11 $GtCO_2eq/yr$ in 2050, adding 249 $GtCO_2eq$ with 0.1°C warming in 2100. Over 90 countries currently have no long-term targets. If they introduce a $CO_2$ net-zero by 2070 (LTS01), this could reduce 2.7 $GtCO_2eq/yr$ in 2050, saving 316 $GtCO_2eq$ and avoiding 0.14°C. EMME, Pakistan and Vietnam can provide the maximum benefit here.

**Table 1.** Enhanced or delayed national climate pledges. All net-zero targets are represented as LTSs for the sake of presentation (many countries include net-zero targets in their NDCs at UNFCCC). *Grey shaded rows represent delayed pledges. #Higher/lower reduction targets refer to either conditional and unconditional targets, respectively, or the upper and lower range of the target, respectively.

| Near-term targets from Nationally Determined Contributions (NDCs) | |
|---|---|
| **NDC01** | Current levels of all higher reduction targets# for 2030 are further increased by 10%. |
| **NDC02*** | Current levels of all lower reduction targets# for 2030 are further decreased by 10%. |
| **Global Methane Pledge (GMP) and additional non-$CO_2$ mitigation targets** | |
| **GMP01** | GMP is additional to the near-term reductions targets declared in NDCs. |
| **GMP02** | All countries continue to reduce $CH_4$ emissions after 2030 to achieve a further 30% reduction by 2100 from 2030 levels. |
| **$CH_4$++** | Countries without GMP reduce $CH_4$ emissions by 30% (relative to 2022) in 2030 and all countries reduce $CH_4$ emissions by 40% (relative to 2022) in 2035. |
| **$N_2O$++** | All countries reduce $N_2O$ emissions by 13% (relative to 2022) in 2030 and by 18% (relative to 2022) in 2035. |
| **Net-zero targets from Long-Term Strategies (LTSs) and NDCs** | |
| **LTS01** | Countries currently having no long-term target reach net-zero in 2070. |
| **LTS02** | Countries currently having a net-zero target after 2050 reach net-zero in 2050. |
| **LTS03** | Countries with a 2050 net zero target year reach net-zero in 2040. |
| **LTS04*** | Countries currently having a net-zero target delay it by 10 years. |
| **LTS05** | Countries currently with a net-zero target follow a net-zero GHG. |
| **LTS06*** | Countries currently with a net-zero target follow a net-zero $CO_2$ only. |
| **Glasgow Leaders' Declaration on Forests and Land Use (LUF)** | |
| **LUF01** | Stopping deforestation with reforestation continuing at current levels + 100% control of forest degradation. |
| **LUF02** | Partial control on current deforestation with reforestation continuing at current levels + 50% control of forest degradation. |
| **LUF03** | Continued deforestation at current levels but with increasing reforestation + 10% control of forest degradation |

It is not clear if GMP should be considered part of or additional to $CO_2eq$ reduction in NDCs. Our baseline is kept conservative, assuming the former. Otherwise (GMP01), it would reduce 0.05°C warming in 2100. In the absence of any specific target, we assume constant $CH_4$ emissions after 2030, while noting that our approach does not take into account multiple gas abatements from a single source (e.g. (57, 58)). While 155 countries have already joined GMP, major $CH_4$ emitters such as Russia, China and India (59) have yet to join. Thus, if all countries reduce 30% of $CH_4$ emissions by 2030 and 40% by 2035 ($CH_4$++) as in the COP28 draft negotiation text, it would further reduce 2 $GtCO_2eq/yr$ in 2050 with a reduction of 166 $GtCO_2eq$ during 2023-2100, avoiding 0.10°C. Additionally, if all countries reduce 13% of $N_2O$ emissions by 2030 and 18% by 2035 ($N_2O$++), it would reduce 37 $GtCO_2eq$ during 2023-2100, avoiding 0.02°C warming.

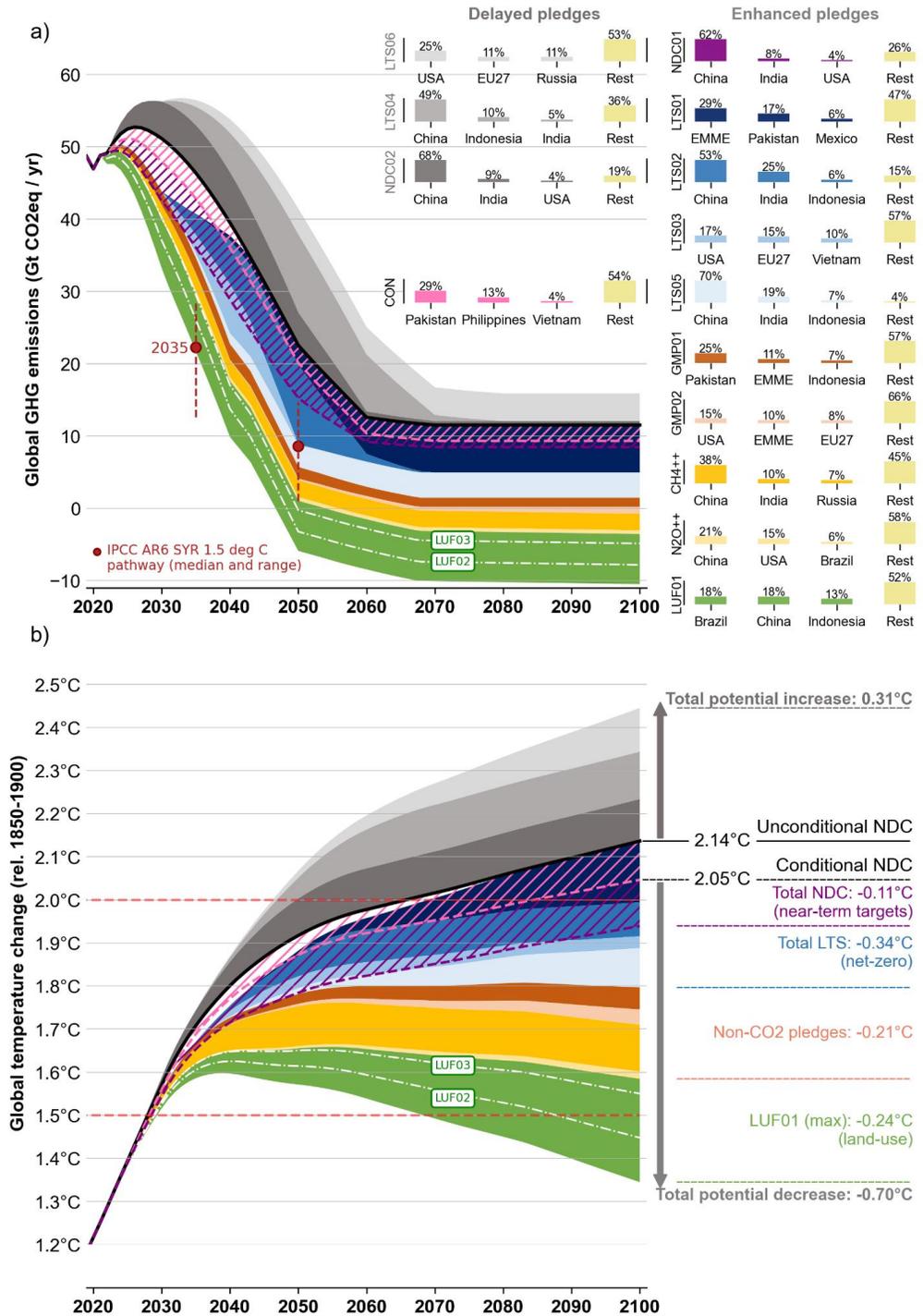

**Fig. 2. Emissions and temperature responses to enhanced or delayed national climate pledges. (a)** Changes to global annual $CO_2$eq emissions per enhanced/delayed pledge (GHG aggregation based on GWP100) and contribution by top three countries to changes in cumulative emissions from 2020-2100 for each enhanced/delayed pledge. **(b)** changes to global temperatures per enhanced/delayed pledge as calculated by the TCRE+GWP+ZEC approach. "CON" indicates the difference between conditional and unconditional targets from countries declaring both. CON and NDC01 pledges (hatched) are shown in a separate layer from other pledges (shaded) due to their potential inter-dependency (as indicated by the overlaps of pledges).

Over 140 countries committed LUF, but implications of this strategy are sensitive to its interpretation (*60*). Among possible interpretations, "end gross forest loss" leads to maximum reduction of $CO_2$ budget, leading to even negative $CO_2$ emissions after 2030. "End tree cover lost" has the least benefit with marginal reductions during the latter half of the century. LUF has the potential to lower warming by 0.03-0.2°C in 2100, depending on its interpretation (see Methods).

Finally, if all these pledges are implemented together, it would reduce 18 $GtCO_2eq/yr$ in 2030 and 34 $GtCO_2eq/yr$ in 2050 relative to current pledges with unconditional targets. This is equivalent to GHG emission reductions of 55% by 2035 relative to 2019 levels, comparable to 60% GHG reductions by 2035 under the IPCC 1.5°C pathway (*61*). Under these most optimistic assumptions, the end-of-century warming would be 1.35°C, and the 1.5°C target could be achieved however still with overshoot.

## Supplemental strategies to comply with the Paris Agreement objectives

Our results highlight that global commitments are currently not aligned with the Paris temperature target. To supplement conventional mitigation efforts, we explore the potential contributions of novel CDR and SRM strategies. Conventional CDR techniques include afforestation, reforestation, management of existing forests and all nature based solutions (*62, 63*). We assume that these would be already accounted for in current pledges, as well as in the enhanced/delayed pledges, which allows us to assess the role of novel CDR separately from pledges. Currently novel CDR constitutes only a negligible removal of 0.002 $GtCO_2/yr$ (*46*). CDR can be counted as removal under the Agreement, despite concerns about uncertainty; however, it is more difficult to justify recognition of SRM under the Agreement because it does not directly affect emissions or removals.

Among various views (*39, 64, 65*), we draw three possible interpretations of the Paris target whereby global temperatures i) peak at 2°C and remain stable henceforth (PA1), ii) peak at 2°C and declining linearly to 1.5°C by 2100 (PA2) and iii) peak at 1.5°C and remain stable henceforth (PA3). These are analogous to the typology of AR6 scenarios: those which keep warming below 2°C (C3), warming below 1.5°C with high overshoot (C2) and without or with low overshoot (C1). For each interpretation, we explore the required levels of novel CDR and SRM to compensate for the remaining gap not met with current pledges (Methods). We prioritize novel CDR; in other words, SRM can be operationalized only if the desired temperature target is infeasible with novel CDR alone, considering CDR levels reported in AR6.

Under current pledges with unconditional targets, complying with the Agreement would require considerable deployment of novel CDR, particularly for 1.5°C (Fig. 3). For PA1, no CDR or SRM is required as the temperature is just below 2°C. For PA2, deployment could start around 2029, removing up to -8 $GtCO_2/yr$ in 2050 and -24 $GtCO_2/yr$ during 2085-2094, following the maximum removal rates reported for C2 scenarios. Finally, for PA3, even with maximum CDR levels for C1 scenarios, CDR alone could not stabilize the temperatures to 1.5°C, especially during 2030-2035. Thus, it would have to be supplemented with SRM at high levels (up to -0.8 $W/m^2$, roughly equivalent to a Mount Pinatubo eruption for every eight years (*66*)) initially, followed by a gradual decrease.

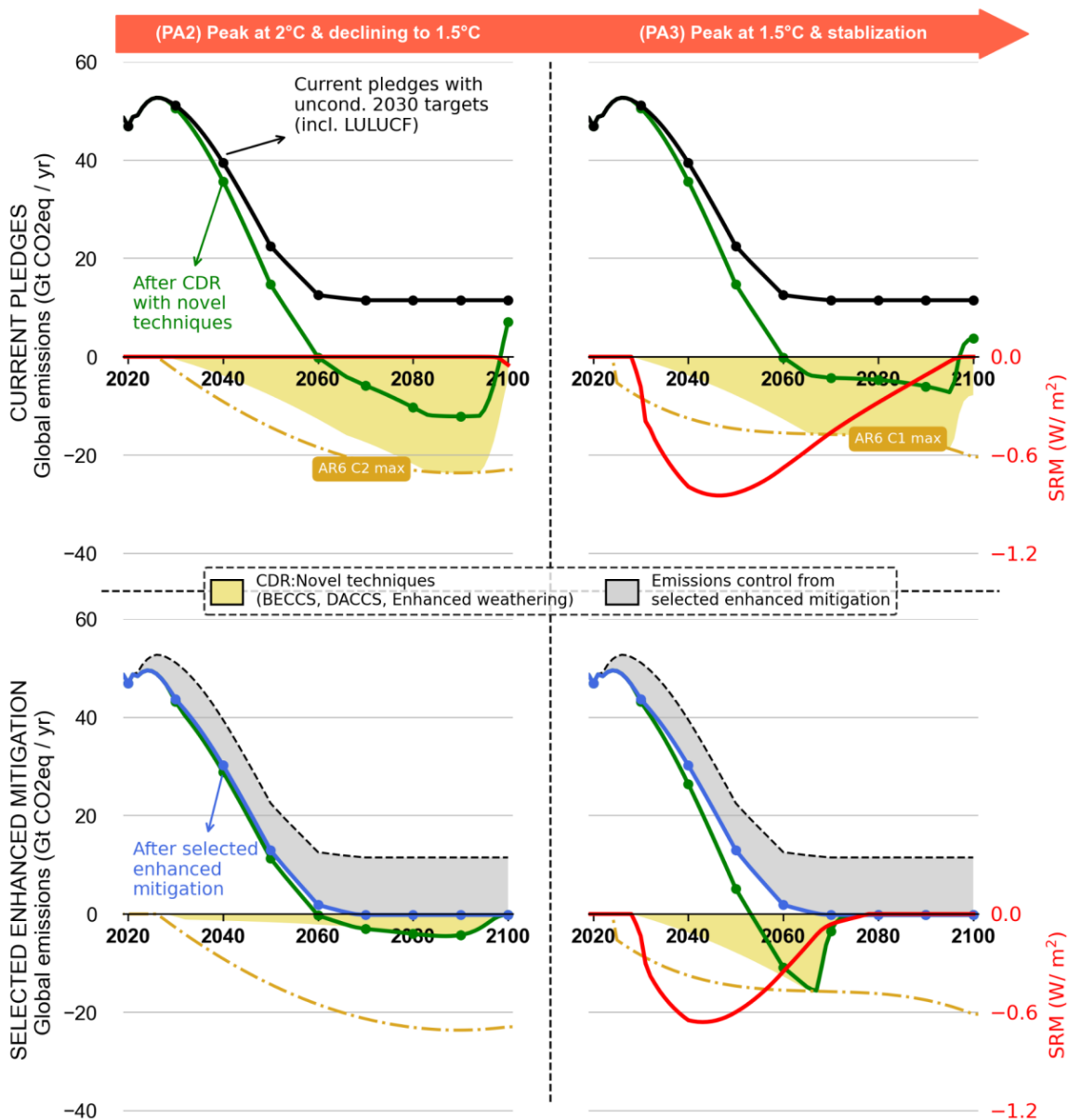

**Fig. 3. Supplementary strategies to comply with the Paris Agreement objectives.** Levels of novel CDR and SRM compatible with different interpretations of the Paris temperature target under current pledges with unconditional targets (top panels) and with selected enhanced pledges (bottom panels). Only the results for PA2 and PA3 are shown since PA1 (peak warming of 1.96 °C) does not require any novel CDR and SRM.

We further explored the extent to which novel CDR and SRM are required with a selected combination of enhanced pledges (conditional targets, GMP01, LTS01 and LUF02). PA2 would be feasible with limited CDR starting from 2029 and reaching a maximum of -4 GtCO₂eq/yr by 2077. However, PA3 would still require SRM, albeit at a lower level. This suggests that, while the selected pledges can reduce the CDR and SRM levels, it cannot totally negate their need, in particular SRM, for the most ambitious interpretation of the Paris objectives. Our analysis further suggests that PA3 requires SRM to begin no later than 2029. Since it may take about 10 years from now to prepare for SRM deployment (*62*), it could be further argued that achieving the 1.5°C target without overshoot is not feasible, even with SRM.

### Recommendations for future national climate pledges

We put forward the following key recommendations that can guide the next revisions of national pledges.

- Commitment to a mid-century net-zero target for all countries, in particular the EMME countries, and an earlier net-zero target for developed countries
- Expansion of the GMP signatories to include countries such as China, India and Russia
- Ambitious implementation of LUF to substantiate emission reductions

Missing near-term targets by 10% or delaying net-zero targets by 10 years could each add 0.1°C to the end-of-century warming, underscoring the importance of implementing pledges once made (*68*). While each enhanced pledge could lower the warming in 2100, it is only through net-zero $CO_2$ or GHG emissions at the global scale and/or increasing land sinks that could effectively halt the warming. Other enhanced pledges are also important but only delay the year of reaching 2°C.

If current pledges are fully realized, novel CDR would be strong enough to keep the end-of-century temperature below 1.5°C. However, achieving the 1.5°C target without overshoot would be impossible without SRM in the near term. In light of this finding, while further evaluation of the SRM scenario is warranted, we call for a reassessment of the interpretation of the 1.5°C target (*69*). The 1.5°C target has been, in our view, crucial in conveying the urgency of climate action; however, a paradigm shift is now necessary to move away from adherence to a hard 1.5°C ceiling to a more flexible interpretation that allows the development of strategies for limiting further overshoot and ultimately reducing the warming towards 1.5°C, accompanied by strategies for adaptation, since every 0.1°C of warming counts. Our study demonstrates unequivocally that it is now unavoidable to exceed 1.5°C. We argue that this message should be clearly and honestly conveyed to policymakers, while emphasizing the importance of reducing the warming to 1.5°C (*70*) and addressing the associated research questions (*71*).

Transparency lies at the core of assessing progress and planning future actions. We have demonstrated a simple, traceable method for calculating emission and temperature projections by country (Figs. S3 to S5), bringing much needed transparency to this ongoing debate. Our approach may serve as a basis for informing the forthcoming revisions of NDCs.

**Acknowledgments:** We are grateful for helpful comments and suggestions from Yann Gaucher, Temuulen Murun, Jean Sciare, Weiwei Xiong, Biqing Zhu, and Eric Zusman. We appreciate the initial help from Clara Thuelin for compiling and processing the NDC files.

**Funding:** This work was conducted as part of the Achieving the Paris Agreement Temperature Targets after Overshoot (PRATO) Project under the Make Our Planet Great Again (MOPGA) Program and benefited from State assistance managed by the National Research Agency in France under the Programme d'Investissements d'Avenir under the reference ANR-19-MPGA-0008. This project has received funding from the European Union's Horizon Europe research and innovation program under Grant Agreement N° 101071247 (Edu4Climate – European Higher Education Institutions Network for Climate and Atmospheric Sciences). K. Tibrewal is supported by the European Union's Horizon 2020 research and innovation program under Grant Agreement N° 856612 (EMME-CARE – Climate change, atmospheric research centre in Eastern Mediterranean and Middle East Region). We further acknowledge the European Union's Horizon Europe research and innovation program under Grant Agreement N° 101056939 (RESCUE – Response of the Earth System to overshoot, Climate neUtrality and negative Emissions) and Grant Agreement N° 820829 (CONSTRAIN – Constraining uncertainty of multi-decadal climate projections).

**Author contributions:** O.B. and K.Ta conceived this study. K.Ti., K.Ta, and O.B. designed the experiment. K.Ti, K.Ta, and O.B. performed the analysis. K.Ti generated the figures and tables. K.Ti., K.Ta, P.C., and O.B. analyzed the results. K.Ti. and K.Ta drafted the manuscript, with contributions from all coauthors.

## Supplementary Materials

Methods, Figs. S1 to S5, Tables S1 to S2, and References (*72–102*)

**Methods**

Emission datasets

To develop future projections of emissions and temperature changes, we have compiled a time series of emissions from multiple datasets. Historical emissions of $CO_2$, $CH_4$ and $N_2O$ from 1990-2022 for major IPCC sectors (Energy, Industrial Processes and Product Use, Agriculture and Waste) are obtained for each country from the latest version of PRIMAP-hist v2.5 (*72*, *73*). Global emissions for international aviation and shipping for 1970-2022 are obtained from EDGARv8 (*74*).

$CO_2$ flux from LULUCF, also referred to as forestry and other land use (FOLU), comprises emissions from deforestation, sinks from forests over managed land and emissions due to forest degradation. Historical $CO_2$ fluxes (1750-2020) for deforestation per country are obtained from the OSCAR bookkeeping model (*34,35*) and sinks over managed land are based on dynamic global vegetation models (DGVMs) (*77*), both of which are part of the Global Carbon Budget 2022 (*76*). Emissions from forest degradation can be quite significant for tropical forests and roughly represent one-third of emissions from deforestation at the global level with even higher contributions for certain countries (*78*). Due to lack of comprehensive data, historical emissions from forest degradation are developed at a global level assuming one-third of total deforestation emissions from the 74 tropical countries listed in ref. (*78*).

Global emissions for other climate forcers including the fluorinated GHGs (chlorofluorocarbons - CFCs, hydrochlorofluorocarbons - HCFCs, hydrofluorocarbons - HFCs, perfluorocarbons - PFCs, sulphur hexafluoride - $SF_6$) and air pollutants and aerosol precursors ($SO_2$, $NO_x$, VOCs, CO, constituents of particulate matter) are assumed from SSP1-1.9 (default) for the period 1990-2100 (*53*).

Interpreting climate pledges

Modeling of future emissions begins by collecting and interpreting information contained in NDCs. Near-term targets usually consist of a $CO_2$ or GHG reduction level in their NDCs. For each country we use the latest version of their NDC (cut-off date: 13 December 2023) (*51*). These are declared in one of the following ways: i) as a reduction in their carbon intensity (i.e. carbon emissions per unit of gross domestic product), ii) with respect to a projected BAU emissions in 2030 or iii) with respect to emissions in a past year (usually 1990, 2005 or 2010, among others). The reduction level is usually expressed as percentage reduction in $CO_2$ only or $CO_2$eq emissions in 2030 with respect to the base (either 2030-BAU or past year) combining all major ($CO_2$, $CH_4$, $N_2O$) and sometimes additional (PFCs, HFCs, $SF_6$, and $NF_3$) GHGs, or so-called Kyoto gases, using GWP100 as a common metric (*29*). Some countries declare two targets, one that can be attained using indigenous resources (unconditional target) and the other with some additional reduction subject to international support (conditional target). Some countries declare a range of emission reduction rather than a single number. Our "higher reduction targets" account for conditional targets and upper ranges of emission reduction targets. Our "lower reduction targets" consider unconditional targets and lower ranges of emission reduction targets. The emissions levels in the base year as well as proposed targets are obtained for individual countries from their official NDCs submitted to the UNFCCC (*51*).

Long-term goals refer to the target year (i.e., 2050, 2060 or 2070) for achieving either net-zero emissions or larger reductions than 2030. Net-zero emissions can be defined in two ways: a) net-zero GHG emissions, which consider emissions from several GHGs (i.e., GHG neutrality) and b) net-zero $CO_2$ emissions, which consider only $CO_2$ (i.e., carbon neutrality). Achieving net-zero GHGs would require "negative" $CO_2$ emissions i.e., a net carbon sink, to compensate for residual non-$CO_2$ emissions from difficult-to-abate sectors (*39*, *64*). We obtain specific net-zero types from the respective LTS submitted by the countries, as well as NDCs, since some countries include long-term goals in NDCs. In the absence of any clear indication, we assume net-zero emissions for only $CO_2$. The declared target years for different countries are mostly collected from the official UNFCCC repository (*52*). Furthermore, we assume that net-zero targets are intended for emissions other than LULUCF, unless explicitly stated otherwise in national pledges. This means that our long-term emission estimates are on the conservative side due to the strict interpretation of net-zero targets.

Information on both near-term and long-term targets for international aviation and maritime transport are obtained from statements released by International Civil Aviation Organisation (ICAO) (*79*) and International Maritime Organisation (IMO) (*80*), respectively. For international aviation, near-term assumes constant emissions from 2019 levels following the carbon neutral growth from 2020 onwards as per ICAO and net-zero in 2050 as per International Air Transport Association. For international shipping, near-term assumes 20-30% reductions by 2030 compared to 2008 levels and net-zero in 2050 following revised targets by IMO in July 2023 (81).

We have collected the list of countries participating in GMP to model $CH_4$ emissions (*11*). These countries commit to reduce their $CH_4$ emissions to achieve a global $CH_4$ reduction of 30% by 2030 compared to 2020 levels. Since the countries participating in GMP contribute less than 60% to global $CH_4$ emissions, achieving GMP would require these countries to reduce their current $CH_4$ emissions by 45%. However, given that it is a highly challenging task, we have assumed a conservative scenario where each GMP participating country achieves at least 30% reduction in $CH_4$ emissions.

In regard to LUF, future $CO_2$ emissions from forest deforestation can vary significantly depending on how countries interpret "halting deforestation". Ref. (*60*) presented three possible interpretations - a) "end gross forest loss" (stopping deforestation with reforestation continuing at current levels), b) "end net forest loss'' (partial control on current deforestation with reforestation continuing at current levels) and c) "end tree cover loss" (continued deforestation at current levels but with increasing reforestation) and compared it against a BAU scenario (BAU_deforestation), extrapolating trends from 2015-2020 until 2030 and maintaining that level afterwards. However, it did not include emissions control from forest degradation. We developed scenarios for future emissions related to forest degradation. We first developed baseline emissions for forest degradation (BAU_degradation) by taking one-third of the total emissions from BAU_deforestation for 74 tropical countries listed in ref. (*78*). With respect to this, we developed three scenarios with 100%, 50% and 10% reductions of degradation emissions for "end gross forest loss," "end net forest loss," and "end tree cover loss," respectively, from 2022 onwards and merged with the deforestation scenarios.

Modeling future emissions excluding LULUCF

After compiling various pledges, future $CO_2$, $CH_4$ and $N_2O$ emissions are first estimated for specific time points (i.e., 2030 and the net-zero target year) and then inter- and extrapolated throughout 2022-2100 for each country separately by using an emission model described below. Our emission model assumes that GHGs include $CO_2$, $CH_4$ and $N_2O$. These are the three most important GHGs, which largely determine future temperature projections. Air pollutants, which can also significantly influence the temperatures, have limited representation in national climate pledges and are assumed to follow SSP1-1.9 in our analysis.

$N_2O$ emissions are kept at 2022 levels from PRIMAP-hist till 2100. For countries following GMP, $CH_4$ emissions in 2030 are obtained by reducing 30% from 2020 levels in PRIMAP-hist. For other nations, $CH_4$ emissions in 2030 are scaled proportionally to the ratio of $CO_2eq$ emissions (excluding LULUCF) and $CH_4$ emissions in the base year. $CH_4$ emissions are linearly interpolated between 2018 and 2030 and then assumed to be constant at 2030 levels till 2100.

$CO_2$ emissions involve a more complex approach. We begin by estimating the emissions in 2030 based on the near-term targets from NDCs. In general, countries report total sectoral emissions (base year and/or BAU 2030) and corresponding reduction targets. There are variations among countries regarding the sectors covered in their respective NDC. Some countries may include the LULUCF sector in their reported emissions and reduction targets. So we first modify the reported emissions (for base year and BAU 2030) and reduction targets such that it does not include the LULUCF sector. For countries that report emissions and reduction targets by each sector, we consider values without the LULUCF sector. For countries that include the LULUCF sector in their NDC but do not provide separate sectoral values for emissions and targets, we generate the emissions without LULUCF (for base year and/or BAU 2030) by subtracting the LULUCF flux compiled by ref. (77) from the total $CO_2eq$ emissions in the NDC. In the latter case, we assume the same values for percentage reduction targets without LULUCF as given for the total sectors. Finally, emissions are combined with the corresponding unconditional and conditional targets to arrive at the $CO_2eq$ emissions without LULUCF in 2030.

Since more than two-thirds of the countries include either/all of the three – $CO_2$, $CH_4$ and $N_2O$ – with unclear or no mention of other GHGs, we distribute the total $CO_2eq$ among these three gases through the method described above. Furthermore, it is assumed that any reductions in $CH_4$ will cater to both NDC target and $CH_4$ pledge (i.e., our base assumption is that $CH_4$ reductions via GMP are also counted as part of progress toward meeting NDC). Thus, 2030 emissions for $N_2O$ and $CH_4$ as calculated above are converted to $CO_2eq$ using GWP100 values in IPCC AR5 as adopted by the Paris Agreement rulebook (265 and 28 for $N_2O$ and $CH_4$, respectively) and subtracted from the total 2030 $CO_2eq$ emissions to obtain the values for $CO_2$ emissions in 2030. This implies that the use of a non-$CO_2$ pledge can result in an increase in $CO_2$ emissions, which can be further sustained under our model assumptions. For example, a scenario with no GMP will give lower $CO_2$ emissions as now more $CO_2$ has to be reduced to meet the desired reductions in total $CO_2eq$ in 2030 and in the neutrality year (if net-zero is for GHG). Note further that for countries where NDC does not provide any base value (primarily for the past year), we assume the value reported in PRIMAP-hist (e.g., India, China, Kazakhstan, Kenya) . For countries, where the base year emissions in PRIMAP-hist are 20% larger than those in NDC, we adjust the BAU 2030 emissions in the same proportion (e.g., Afghanistan, Angola, Lao, Tunisia).

For countries without any long-term target, emissions are assumed to be constant at the 2030 level up until 2100. For countries having a neutrality target, we either assign zero or negative $CO_2$ emissions compensating for $CH_4$ and $N_2O$ emissions in the proposed year, based on their net-zero type. Then using emissions in 2022, 2030 and the corresponding long-term target year for $CO_2$ (Fig. S3), we fit a curve based on an exponential function (Eq. 1) from ref. (*17*),

$$E_y = E_{asm} + (E_{2022} - E_{asm}) \times e^{g_y} \quad \dots (1)$$

where,

$E_y$ = $CO_2$ emissions in year $y$ (from 2023 to 2100)

$E_{2022}$ = $CO_2$ emissions in year 2022 from PRIMAP-Hist

$E_{asm}$ = Asymptotic $CO_2$ emissions at the final target level

$g_y$ = Cumulative growth rate $\int_{2023}^{y} g(t)dt$

$g(t)$ = Annual growth rate $g_t = g_{t-1} - \Delta g$

$\Delta g$ = Change in the annual growth rate

The values of $g_{2022}$ are obtained from the past trends 2018-2022 while those of $E_{asm}$ and $\Delta g$ are obtained by optimizing the function to minimize the misfit from the emissions in 2030. Emissions are assumed to be constant beyond the long-term target year.

Our emission pathways are more conservative than those in previous studies because we do not assume further emission reductions beyond the target level (i.e., Constant Emissions approach (*18*)). For countries with a net-zero GHG target, net negative $CO_2$ emissions can occur to a certain extent to compensate for residual non-$CO_2$ emissions from hard-to-abate sectors, but no further.

Modeling future emissions for LULUCF

The modeling for the LULUCF sector starts from 2021 to 2100, as the historical dataset is available up to 2020. LULUCF comprises anthropogenic $CO_2$ fluxes from three components - i) emissions from deforestation, ii) sinks induced from environmental change over managed land and iii) emissions from forest degradation. For (i) and (ii), we performed a country-level analysis, while (iii) is accounted for at a global level.

For top contributors to the net fluxes from i) and ii) combined (*77*), we obtain the net fluxes in 2030 either directly from the NDC report when stated explicitly or from the interpretation by ref. (*81*). For the rest of the countries, the net fluxes from i) and ii) are kept constant at 2020 levels from historical datasets. Since LULUCF fluxes reported in NDC or ref. (*81*) only include i) and ii), we add global emissions for iii) from ref. (*78*) as average emissions during 2005-2010.

All temperature simulating approaches used here internally calculate total $CO_2$ uptake over land, including the total sinks over both managed and unmanaged land. In order to avoid double counting of sinks from managed land (Fig. S1), we exclude the sinks from the total emissions to represent only the direct LULUCF fluxes (i.e., emissions from deforestation and degradation). In this regard, the following steps were taken:

- Step1: ACC2's land uptake parameterization (those representing $CO_2$ fertilization (beta) and climate-carbon cycle feedback (Q10)) is tuned such that the sum of deforestation emissions from OSCAR and total $CO_2$ uptake over land from ACC2 is within the levels of net $CO_2$ flux over land as computed from other several combinations; OSCAR deforestation plus total sinks from DGVMS (managed and unmanaged); net flux from NGHGI data (i.e., deforestation emissions and sinks over managed land) plus additional sinks over unmanaged land based on inversions; total net flux from inversions (Fig. S2). The values of beta and Q10 were set at 0.6 and 1.2, respectively.

- Step 2: For the historical period (up to 2020), double counting is avoided by merely excluding the DGVM sinks over managed land in the emission input, i.e., by considering only the emissions from deforestation and degradation.

- Step 3: For the modeled future LULUCF flux (2021-2100), the sink component over managed land is difficult to disaggregate due to lack of information. Thus, we subtract sinks assuming a linear decrease in land sinks from 2020 levels (as reported by DGVMs) to zero in 2100, following the decreasing sinks simulated by many models under low emissions scenarios (*82*, *83*), as a representative approximation of future share from sinks over managed land.

We incorporated the impacts from LUF at the global level as shown in Table S1.

Simulating temperature change

To ensure robustness, we estimate future temperature projections using three different approaches, representing varying levels of complexity in simulating the response of the climate system to changing emissions as described in the order of relatively high to low complexity. We use the third approach as the representative method for temperature projections of enhanced/delayed pledges (Fig. 2), as well as those of current pledges (Fig. 1). The temperature projections of current pledges are compared with those based on the first and second approaches (Fig. 1). The first approach is also used in the analysis of supplementary strategies that explore the potential use of novel CDR and SRM (Fig. 3).

In the first approach, temperature responses are simulated using a reduced-complexity climate model ACC2 (*38–40*). The current model was developed from earlier reduced-complexity climate models (*84*, *85*). The model comprises four modules: carbon cycle, atmospheric chemistry, climate, and economy modules. The economy module enables ACC2 to calculate least-cost pathways (*39*, *40*, *86*) for a given mitigation target. However, this study uses ACC2 without the economy module to simulate only the climate aspects. The performance of this model was cross-compared with those of other reduced-complexity climate models (*82*, *87*, *88*). The model calculates the temperature contributions of various gases and aerosols separately ($CO_2$, $CH_4$, $N_2O$, 29 species of halocarbons, $SF_6$, tropospheric and stratospheric $O_3$, stratospheric water vapor, and aerosols (the direct effect of sulfate aerosols, the direct effect of black carbon and organic aerosols, and the indirect effects of all aerosols)) by evaluating the radiative forcing of climate forcers individually (i.e., without any gas aggregation using emission metrics such as GWP100). The physical climate module is an energy balance and heat diffusion model DOECLIM (*38*, *89*). The atmospheric chemistry module takes into account the OH chemistry to compute $CH_4$ and tropospheric $O_3$ concentrations, considering pollutant emissions ($NO_x$, CO, and VOC). The carbon cycle module is a box model comprising three ocean boxes, a coupled atmosphere-mixed layer box, and four land boxes. With rising atmospheric $CO_2$ concentrations, the ocean $CO_2$ uptake can be saturated through changes in the thermodynamic equilibrium of carbonate species, and the land $CO_2$ uptake can logarithmically increase due to the $CO_2$ fertilization effect (beta parameterization). Climate-carbon cycle feedback is considered using a Q10 parameterization. Equilibrium climate sensitivity is one of the major uncertain parameters that determine global average temperature changes in model calculations, with ranges of 1.5-4.5°C in AR5 and 2.5–4.0°C in AR6 for a doubling of $CO_2$. Following the best estimate of AR6, we assume a climate sensitivity of 3°C. Other uncertain model parameters are calibrated based on an inversion approach since 1750 (*90*). The model is written in GAMS and numerically solved using CONOPT3, a nonlinear optimization solver included in the GAMS software package.

The second approach involves a less complex method based on IRF, a simple mathematical model, usually in the form of a sum of exponential functions, to represent a time-dependent response of a perturbation in a system. We combine two different IRFs obtained from Section 8.SM.11 of IPCC AR5 (*41*): i) a carbon cycle IRF that calculates atmospheric $CO_2$ concentrations for a given pulse emission of $CO_2$ and ii) a climate IRF that calculates global-mean temperature changes for a given pulse input of radiative forcing. To calculate temperatures for $CO_2$ emissions in time series, we use double convolutions for these two IRFs. In the temperature calculations, the $CO_2$ forcing is linearly related to the $CO_2$ concentration through the $CO_2$ radiative efficiency in AR5.

There are variations in this calculation method, depending on the way how non-$CO_2$ components are treated (Table S2). In the first case (m1), both $CH_4$ and $N_2O$ concentrations are calculated based on the respective simple gas cycle models, which were used for computing emission metric values in AR5 (*41*). The additional $CH_4$ forcing, or so-called indirect effect, such as the production of tropospheric $O_3$ and stratospheric water vapor is considered as a fixed fraction of the radiative forcing of $CH_4$ itself (direct effect) computed from the $CH_4$ gas cycle model. $CO_2$ production through $CH_4$ oxidation (*91*) is not accounted for in AR5 metric values. The $CH_4$ and $N_2O$ concentrations are converted to $CH_4$ and $N_2O$ forcings via their respective radiative efficiencies, which become part of the total forcing, the input for temperature calculations. In all other cases (m2 to m6), the contributions of $CH_4$ and $N_2O$ are not explicitly considered through models but represented as part of the $CO_2$eq emissions using emission metrics (AR5 metric values, following the Paris Agreement rulebook (*29*)). Metric values in AR6 (*92*) are not used in our analysis despite their higher scientific robustness. The last case (m6) uses the Global Warming Potential Star (GWP*) approach (*21–25*), a metric, or more precisely a revised way of using metrics, that allows emissions of short-lived climate forcers such as $CH_4$ to be converted to $CO_2$eq emissions while maintaining temperature impacts as consistent as possible. The GWP* approach requires the use of $CH_4$ emissions 20 years in the past to quantify the rate of change in $CH_4$ emissions. Finally, the temperature contributions of gases and pollutants other than $CO_2$, $CH_4$ and $N_2O$ ($\Delta T_{other-gases}$) are based on the data for SSP1-1.9 (Fig. 6.22 of IPCC AR6 WGI (*65*)). The temperature change from the IRF+metric approach ($\Delta T^{IRF+metric}$) relative to 1850-1900 levels is obtained from Eq. 2.

$$\Delta T^{IRF+metric}{}_{yr} = (T^{IRF+metric}{}_{yr} - T^{IRF+metric}{}_{2019}) + \Delta T^{Obs}{}_{2019} + \Delta T_{other-gases,yr}$$

$$\dots (2)$$

where,

$yr$ = Years from 2020, up to 2100

$T^{IRF+metric}{}_{yr}$ = Temperature response for $yr$ (relative to 1750) from $CO_2$, $CH_4$ and $N_2O$ emissions calculated using IRF in combination of respective metrics as described as cases in Table S2

$\Delta T^{Obs}{}_{2019}$ = Temperature change in 2019 (relative to 1850-1900) based on observations (= 1.183°C). It is obtained using the decadal mean temperature change (2013-2022) of 1.15°C (representative of temperature change in 2017-2018) and interpolating to 2020 using a decadal rate of warming of 0.22°C/decade (i.e., 1.15°C + 0.22°C/decade * (1.5/10) = 1.183°C). The values are obtained from ref. (*94*).

$\Delta T_{other-gases,yr}$ = Temperature change for $yr$ (relative to 1850-1900) from the emissions of gases and pollutants other than $CO_2$, $CH_4$ and $N_2O$ following SSP1-1.9. This is taken for total aerosols, HFCs and ozone from Fig 6.22 of IPCC AR6 WGI (*93*).

The third approach is the simplest, involving the use of a linear relation between the peak temperature change and the cumulative $CO_2$ emissions in the form of TCRE. It is the ratio of the global mean temperature change per unit $CO_2$ emitted. As per AR6, TCRE ranges from 0.27°C to 0.63°C per 1,000 GtCO₂ with the best estimate of 0.45°C per 1,000 GtCO₂ (*34*). Using the best estimate for TCRE, the global mean temperature change is estimated by Eq. 3. We calculate the temperature change from the 2019 level $\Delta T^{Obs}{}_{2019}$.

ZEC has been known as an additional warming or cooling for certain periods after the point of zero emissions. However, it was recently shown that the effect of ZEC already occurs from the point of the peak $CO_2$ emissions (long before the point of net zero $CO_2$ emissions) (*35, 96, 97*), which may be related to the response of the ocean carbon cycle (*83*). The currently available estimates of ZEC from many models are derived using idealistic scenarios (*36*). The most common indicator of ZEC is ZEC50 (*37, 98*), the temperature change following zero emissions for 50 years, after an exponential increase of $CO_2$ concentrations by 1% per year starting from 1850 (i.e., doubling $CO_2$ concentrations after 70 years) (*99*). Although further research is required to better understand how to incorporate the ZEC contribution in transient temperature calculations such as ours, we make an attempt to incorporate ZEC by assuming that its influence will gradually and linearly come into force in the form of ZEC50 from the point of peak GHG emissions (2022, GWP* basis) up to 2100.

Regarding the non-$CO_2$ contribution, we use the GWP* approach (see above) for $CH_4$ emissions and GWP100 for $N_2O$ emissions to calculate the total $CO_2$eq emissions used as the input for TCRE. We also apply this TCRE+GWP+ZEC method to perform a simple apportionment of the future warming levels at the country level according to modeled $CO_2$eq emissions for each country (Eq. 4), while we acknowledge that this simple approach does not consider associated nonlinearities (*55, 100–102*).

$$\Delta T^{TCRE+GWP+ZEC}{}_{yr} = \Delta T^{Obs}{}_{2019} + \Delta T_{other-gases,yr} + \sum_{t=2020}^{yr} E_t \times TCRE + (a_t \times ZEC_{50})$$

$$\dots (3)$$

where,

$E_t$ = $CO_2$eq emissions with LULUCF (adjusted to remove sinks over managed land; Fig. S1) at year $t$. $CH_4$ and $N_2O$ emissions are converted to $CO_2$eq emissions using GWP* and GWP100, respectively.

$a_t$ = a factor linearly increasing from 0 in year of peak GHG emissions (2022, GWP* basis) to 1 in 2100.

$ZEC_{50}$ = -0.079°C, the mean estimate of ZEC50 from Earth system models (ESMs) and Earth system models of intermediate complexity (EMICs) in IPCC AR6 WGI (*98*).

Exploring supplementary novel CDR and SRM approaches

CDR involves removal of $CO_2$ from the atmosphere, effectively reducing its concentration and its impact on the climate. Besides afforestation and reforestation, as well as all nature-based solutions (*62*, *63*), other techniques, termed novel CDRs in this study, have been proposed, including BECCS, DACCS, and enhanced weathering (*47*), although current levels of implementation are very limited (*46*). SRM involves deliberately altering the Earth's albedo (reflectivity) or the amount of sunlight reaching the surface in order to cool the planet. One proposed SRM method is stratospheric aerosol injection, whereby tiny reflective particles are released into the stratosphere to scatter sunlight away from the Earth (*49*). Novel CDR and SRM are usually not explicitly considered in national pledges. Thus, our analysis treats novel CDR and SRM separately from national pledges and explores the role of novel CDR and SRM additional to national pledges.

We used ACC2 to estimate the minimum levels of such intervention that would be needed to supplement national pledges in order to meet the Paris temperature targets. Temperature limits were imposed in three different ways reflecting possible interpretations of the Paris Agreement temperature target as described in the main text. We estimated the minimum levels of novel CDR (and SRM if novel CDR is insufficient) by optimization based on the following objective function (Eq. 4).

$$\text{Minimize: } \sum_{t=2025}^{2100} w_{CDR} \cdot (CDR_t)^p + w_{SRM} \cdot (SRM_t)^p \ldots \text{(4)}$$

- $CDR_t$ denotes the novel CDR levels for year $t$, estimated in terms of negative $CO_2$ emissions per year (GtC/year in the model calculation). CDR levels are constrained by the maximum annual CDR levels from novel techniques (BECCS, DACCS and enhanced weathering) found in the Paris-compliant AR6 scenarios in the categories of C1, C2, and C3 (for PA3, PA2, and PA1, respectively) and by the maximum rate of change found in the AR6 scenarios in all categories (for all PAs). It is assumed that the novel CDR can start in 2029 if required. $w_{CDR}$ is an arbitrary weighting factor for the CDR term (1/(GtC/year)) and fixed at 1.

- $SRM_t$ denotes the SRM levels for year $t$, estimated in terms of negative radiative forcing (W m$^{-2}$) with the maximum allowable level of -2 W m$^{-2}$. Only when the temperature target is infeasible with novel CDR, the SRM option is invoked. This is done through an arbitrary weighting factor for SRM $w_{SRM}$ (1/(W m$^{-2}$)), which is fixed at 50. The value of this weighting factor was chosen such that $SRM_t$ takes values only when the temperature target is not feasible with CDR alone. It is assumed that SRM is allowed to start from 2029 in our model calculations (this is the latest start year required for this analysis).

- For both CDR and SRM terms, we apply a power of 1.2 as $p$, an arbitrary coefficient to avoid numerical fluctuations in the estimated CDR and SRM levels. With a power of 1.0, the estimated CDR levels are similar to the default results but with numerical fluctuations (fluctuations cannot be penalized with a power of 1.0). With a power of 1.5, while the estimated CDR levels also do not have fluctuations, the CDR profile becomes flatter because the squared power penalizes changes in the intervention levels over time.

Comparison with other studies

Compared to the Meinshausen study ($8$) (thereafter, M22), our NDC cut-off date is two years later. Second, ACC2 tends to give a stronger warming than the model of M22 ($82$) due to parameter and structural differences (despite a climate sensitivity of approximately 3°C in both studies). Third, while we do not assume further mitigation without a deeper target, post-net-zero emission pathways of M22 can decline further, which is driven by, in our understanding, IAMs indicating deep negative $CO_2$ emissions. Fourth, SSP1-1.9 has lower $SO_2$ emissions than those in SSP5 (M22). Fifth, our LULUCF emissions are higher due to i) higher deforestation emissions from OSCAR compared to the NGHGI database (M22) and ii) inclusion of forest degradation emissions (Fig. S1).

The latest Emissions GAP report ($1$) (thereafter, U23) includes the near-term unconditional pledges reported till 25 September 2023 and conditional pledges reported till November 2022. Second, compared to this study, U23 projects a higher emissions pathway for the unconditional pledges including net-zero targets because of its underlying strict criteria for selecting net-zero pledges based on ($3$). This leads to a lesser number of countries (seven) meeting their net-zero as compared to this study, wherein all countries with a long-term target are assumed to achieve their respective net-zero emissions. Therefore, the end-of-century temperature rise for unconditional plus net-zero pledges from U23 is higher than this study. Third, for countries without any long-term target, we do not assume any further reduction beyond 2030 levels, but U23 projections assume a continuation of efforts at a similar level of ambition based on carbon prices computed through five IAMs. Fourth, temperature projections are simulated using the Finite Amplitude Impulse Response (FaIR) reduced complexity climate model.

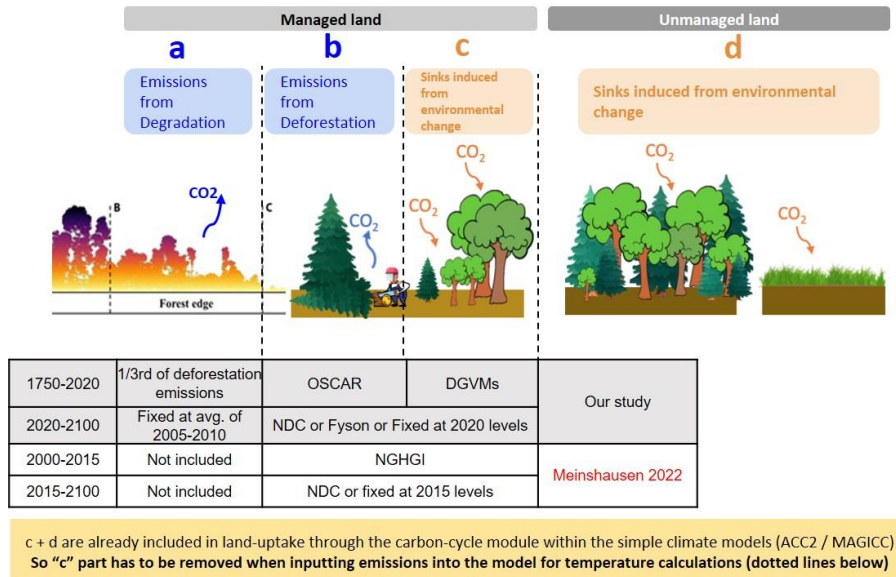

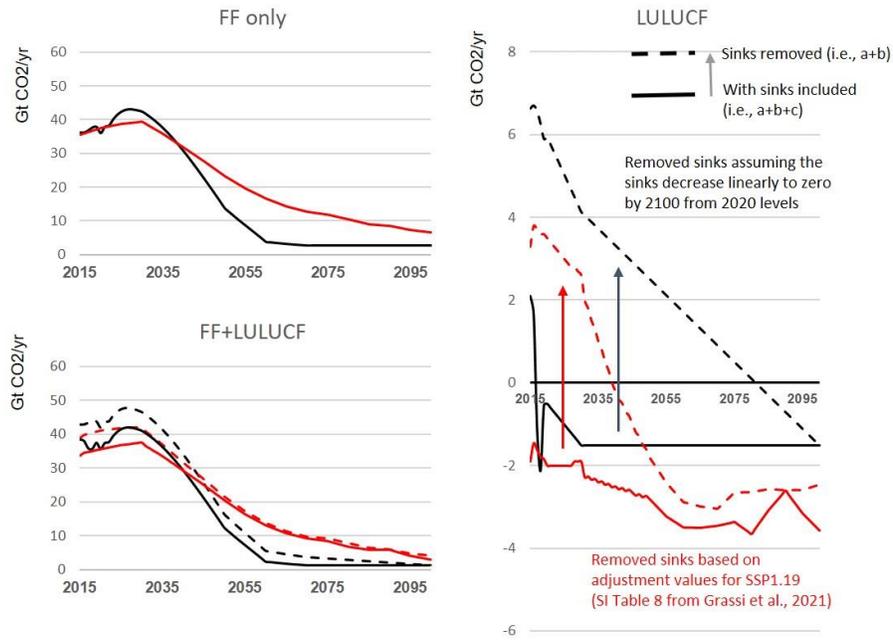

**Fig. S1. Graphical representation of accounting the LULUCF sector.** LULUCF fluxes reported in NDCs or national inventories submitted to UNFCCC contain sinks over managed land. Sinks over managed land must be removed from the emission input to reduced-complexity climate model ACC2, IRF, and TCRE), as they are already included as part of their processes or parameterizations. This graphical representation explains the various components involved in the LULUCF fluxes and how they are accounted for in this study (black line) and Meinshausen et al., 2022 (*8*) (red line). "FF" means $CO_2$ emissions excluding the LULUCF sector, i.e., energy and industrial emissions. The adjustment to LULUCF flux is shown in the right panel. Total $CO_2$ emissions after and before adjusting for sinks are shown on the bottom left panel. The dotted lines in the bottom left panel are input to the respective temperature calculation. For the LULUCF sector, emissions (avg 2015-2020) from net deforestation corresponds to 4 $GtCO_2$/yr and from forest degradation is 2.4 $GtCO_2$/yr (based on 2.1 Gt $CO_2$/yr during 2005-2010 from ref. (*78*)).

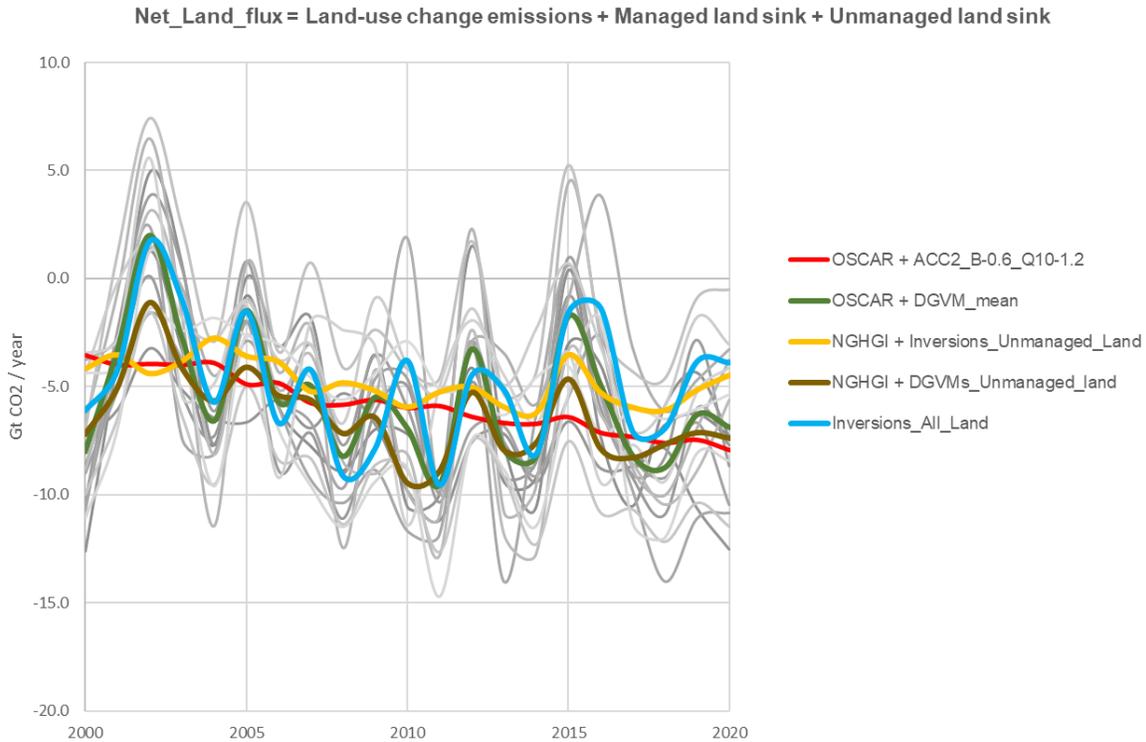

**Fig. S2. Net-land flux from the LULUCF sector using different combinations for data sources.** "Net-land flux" shown here, includes emissions from deforestation and sinks induced from environmental changes over both managed and unmanaged land. "OSCAR" represents emissions from deforestation as reported by the OSCAR bookkeeping model to the Global Carbon Budget 2022. "DGVM_mean" represents the average of total sinks (managed + unmanaged land) from the 16 Dynamic Global Vegetation Models part of the Global Carbon Budget 2022. Shaded grey line represents the sum of OSCAR deforestation emissions and sink from each DGVM. "NGHGI" represents the net-flux over managed land as reported in the national greenhouse gas inventories of countries, submitted to UNFCCC as part of their NIR/BUR/National communications. Inversions represent land fluxes as computed from atmospheric inversions for Global Carbon Budget 2022.

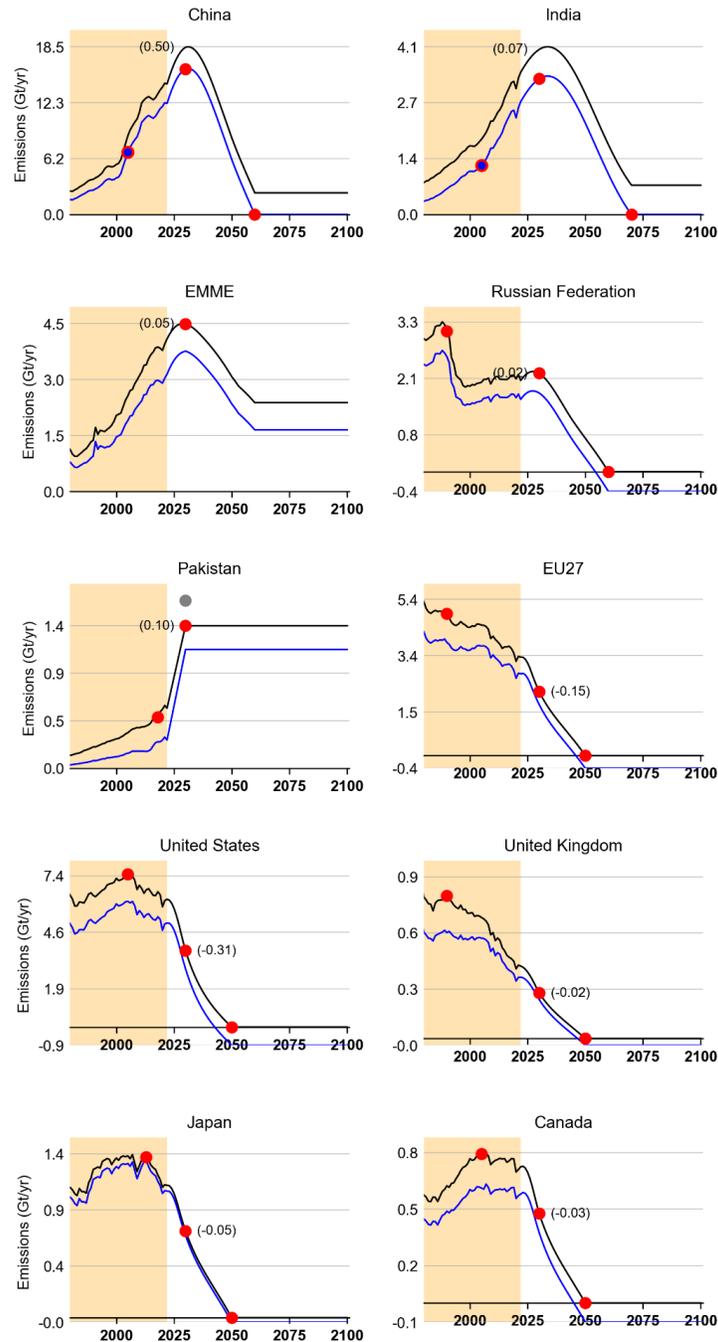

**Fig. S3a. Country-level emissions trajectories for major emitters from Fig. 1.** Emissions trajectories (excluding LULUCF) from 1980-2100 are shown for selected major emitters for $CO_2eq$ (black line) and $CO_2$ (blue line). Dots represent emissions in the base year, BAU 2030, 2030 after reduction, and net-zero (if any), respectively. Generally red dots represent values taken directly from NDC or after applying reductions based on NDC. If the base year is not present in NDC, it is taken from PRIMAP and shown with red-edge black circle ($CO_2eq$) or red-edge blue circle ($CO_2$). BAU 2030 taken from NDC is shown in grey. If BAU 2030 is adjusted based on emissions in the base year, it is shown as grey-edge circle. The value in the bracket denotes the trend in emissions ($Gt/yr^2$) from 2022 to 2030.

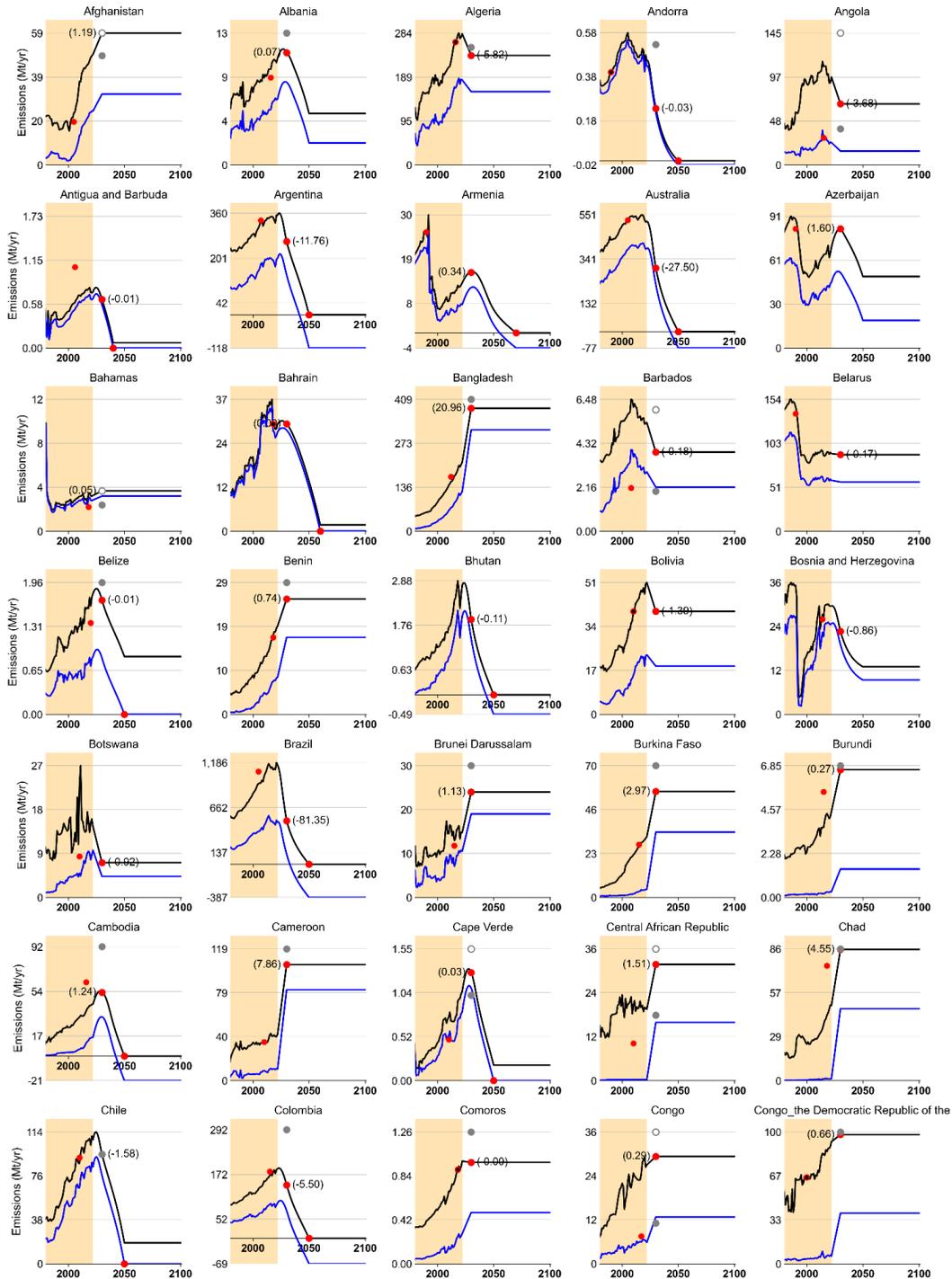

**Fig. S3b. Country-level emissions trajectories.** Emissions trajectories (excluding LULUCF) from 1980-2100 are shown for selected major emitters for $CO_2eq$ (black line) and $CO_2$ (blue line). Dots represent emissions in the base year, BAU 2030, 2030 after reduction, and net-zero (if any), respectively. Generally red dots represent values taken directly from NDC or after applying reductions based on NDC. If the base year is not present in NDC, it is taken from PRIMAP and shown with red-edge circle. BAU 2030 taken from NDC is shown in grey. If BAU 2030 is adjusted based on emissions in the base year, it is shown as grey-edge circle. The value in the bracket denotes the trend in emissions ($Gt/yr^2$) from 2022 to 2030.

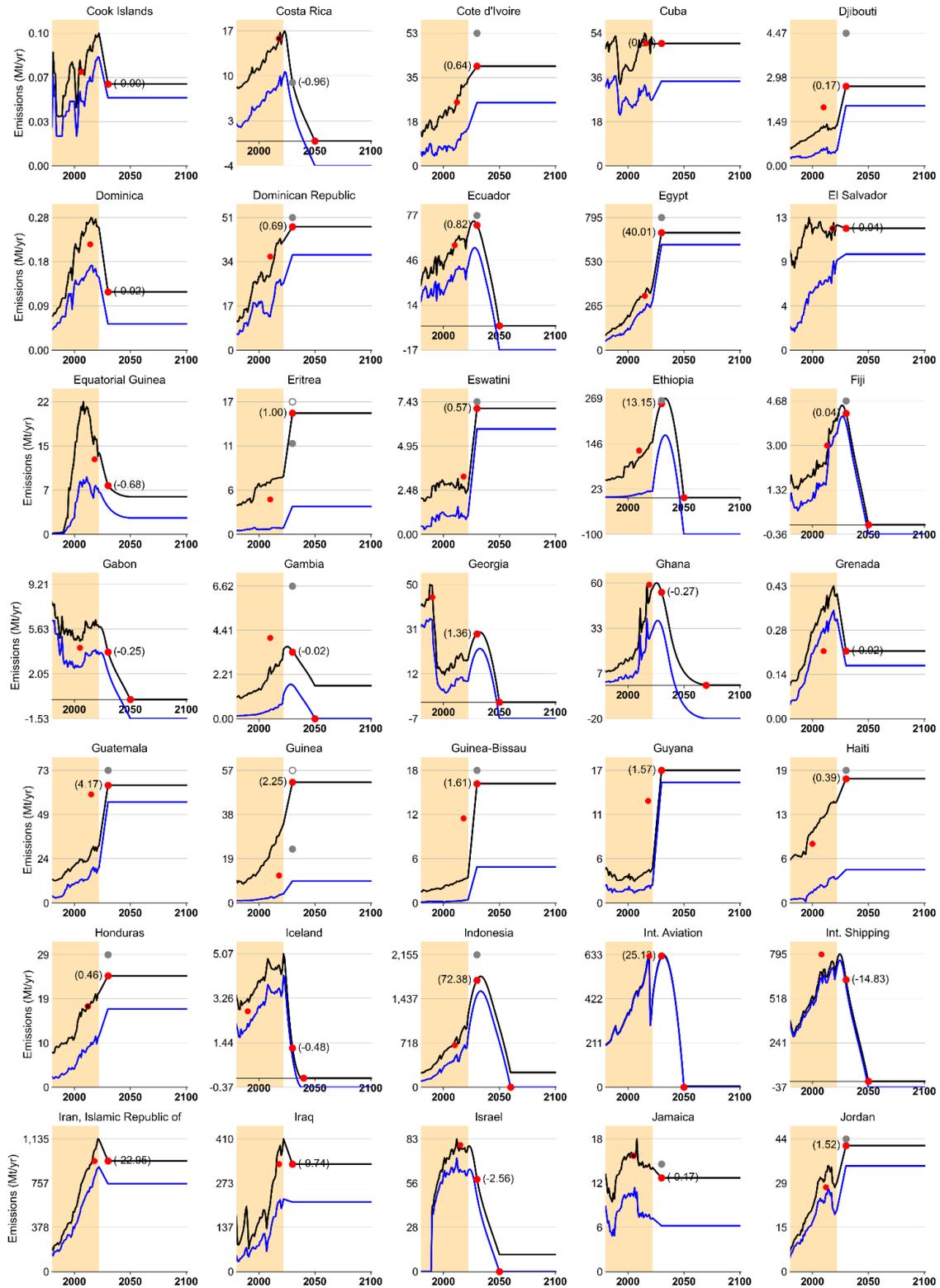

**Fig. S3c. Country-level emissions trajectories.** Same as Fig. S3a but for other countries by alphabetical order.

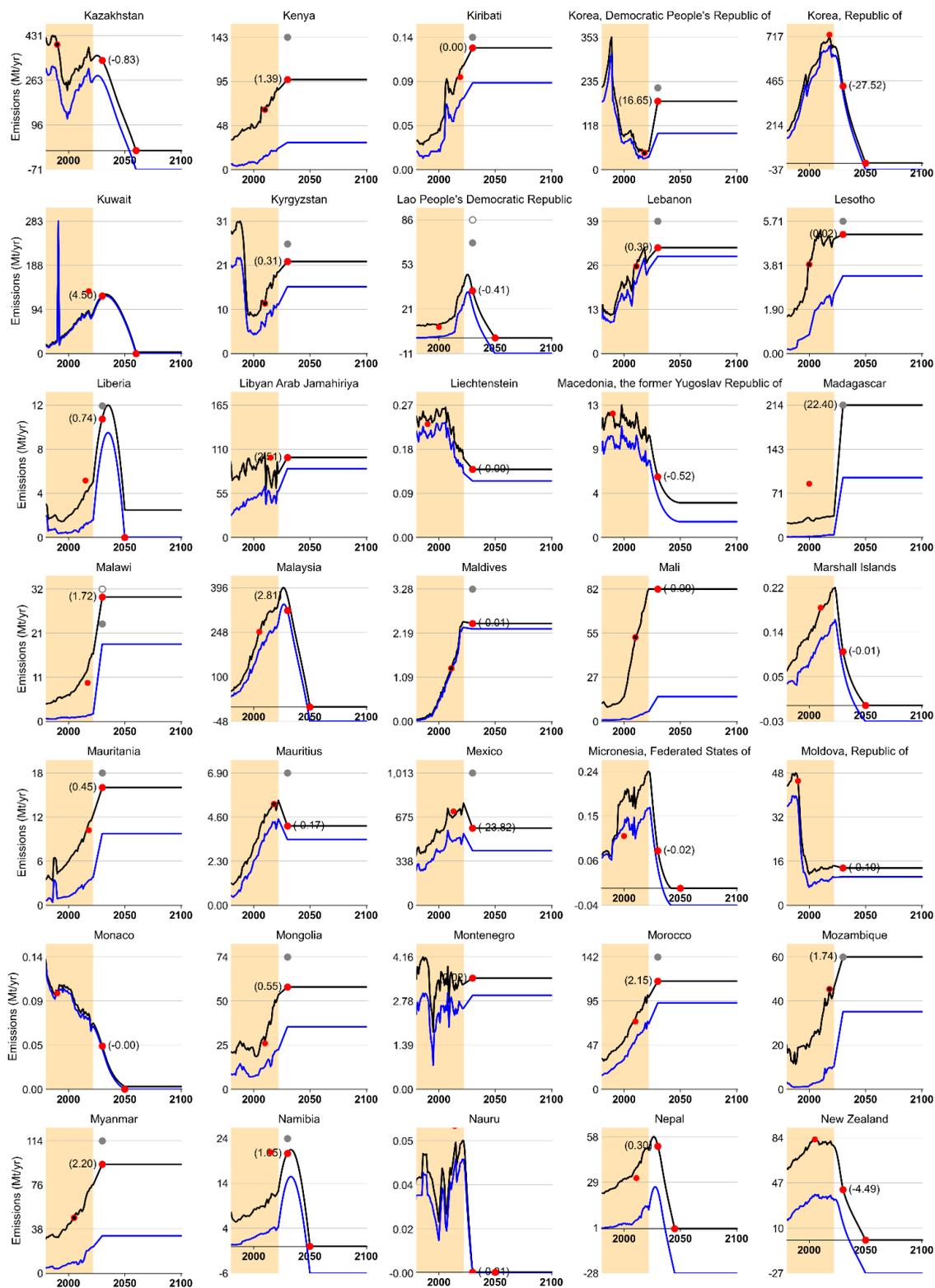

**Fig. S3d. Country-level emissions trajectories.** Same as Fig. S3a but for other countries by alphabetical order.

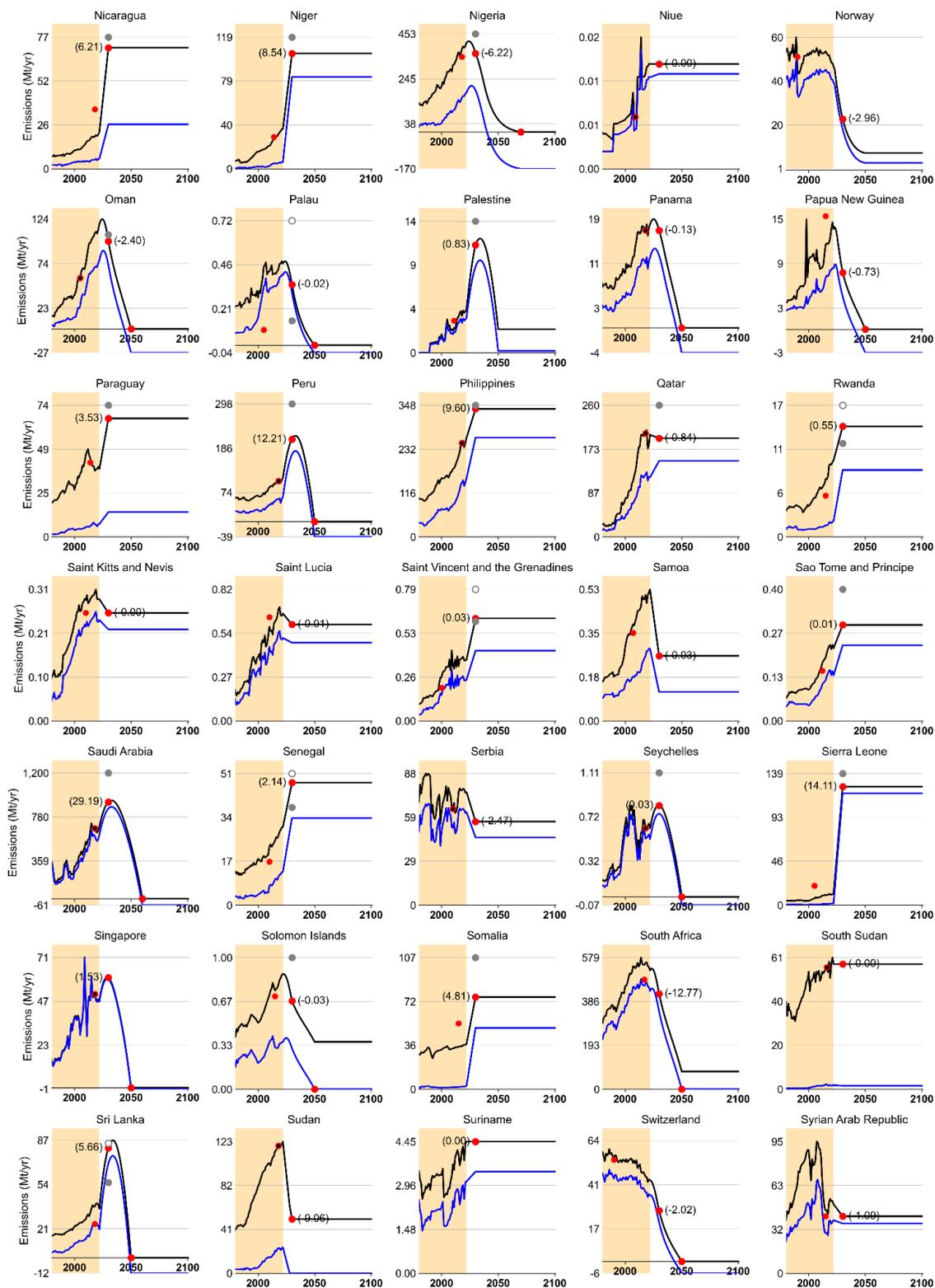

**Fig. S3e. Country-level emissions trajectories.** Same as Fig. S3a but for other countries by alphabetical order.

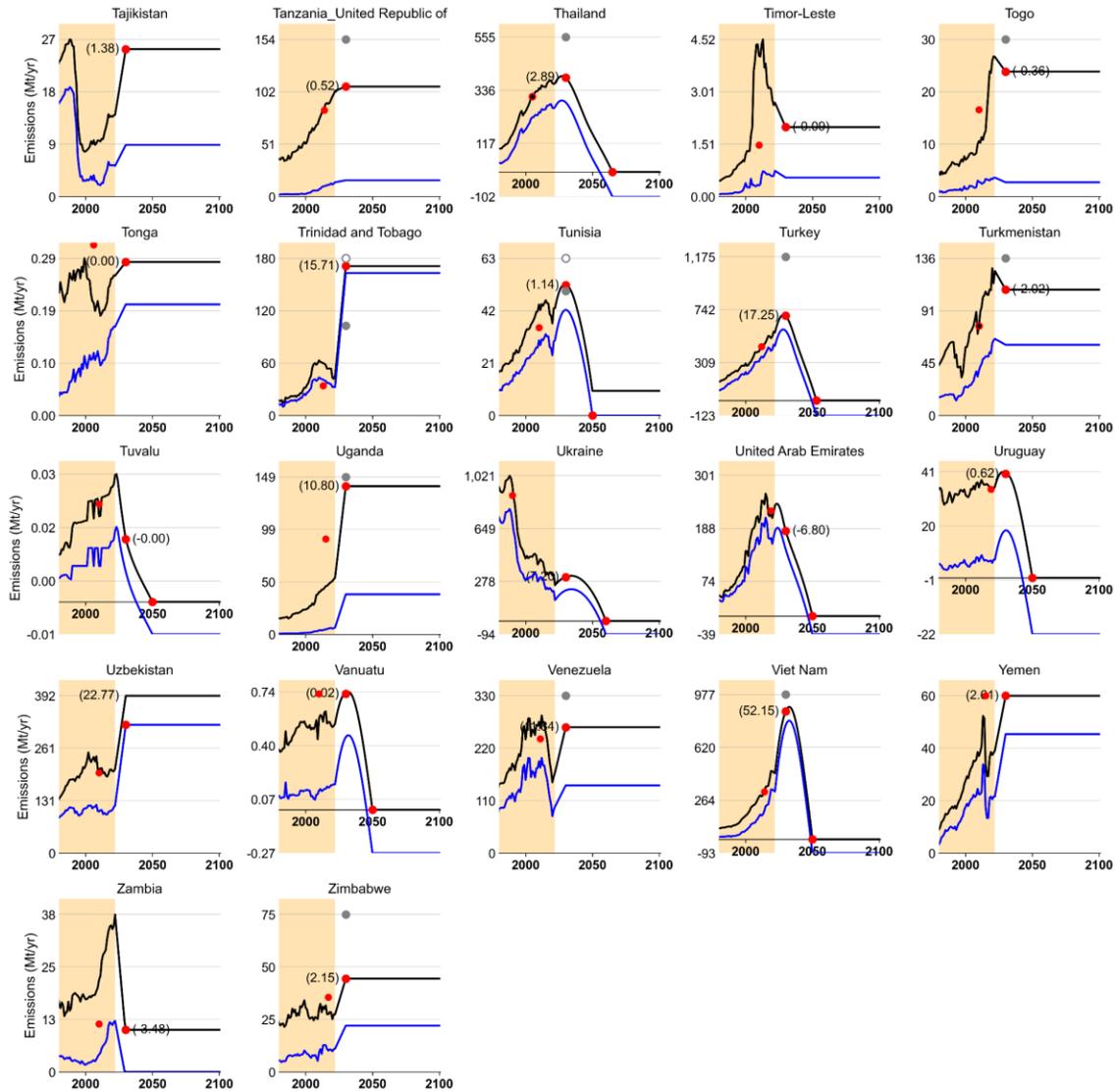

**Fig. S3f. Country-level emissions trajectories.** Same as Fig. S3a but for other countries by alphabetical order.

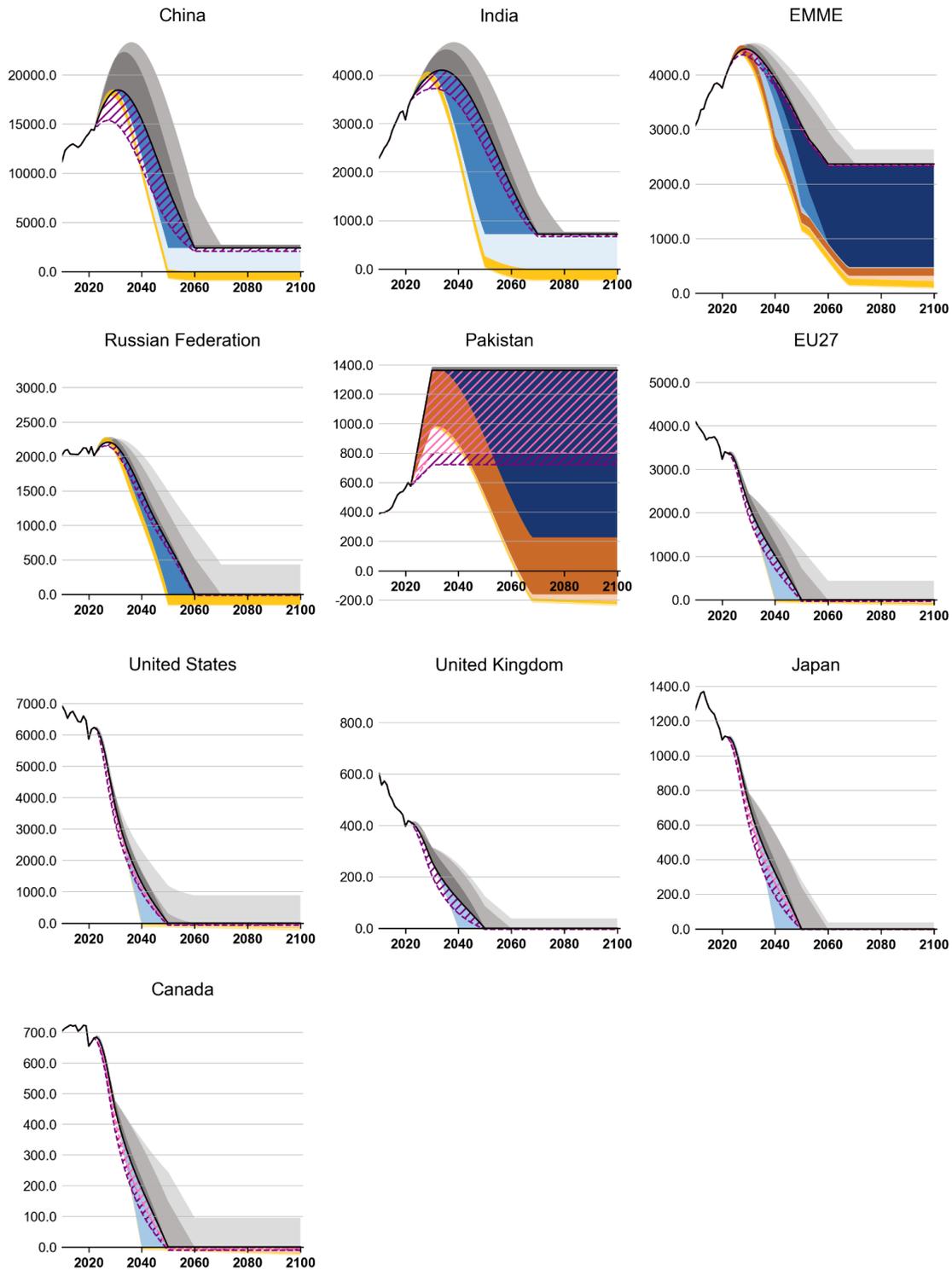

**Fig. S4a. Emissions responses to enhanced or delayed national climate pledges.** CO$_2$eq emissions (Gt/yr) excluding the LULUCF sector are shown for major emitters from Fig. 1. Responses from LUF pledges are not shown as they are estimated only at the global level. CON and NDC01 pledges (hatched) are shown in a separate layer from other pledges (shaded) due to their potential inter-dependency.

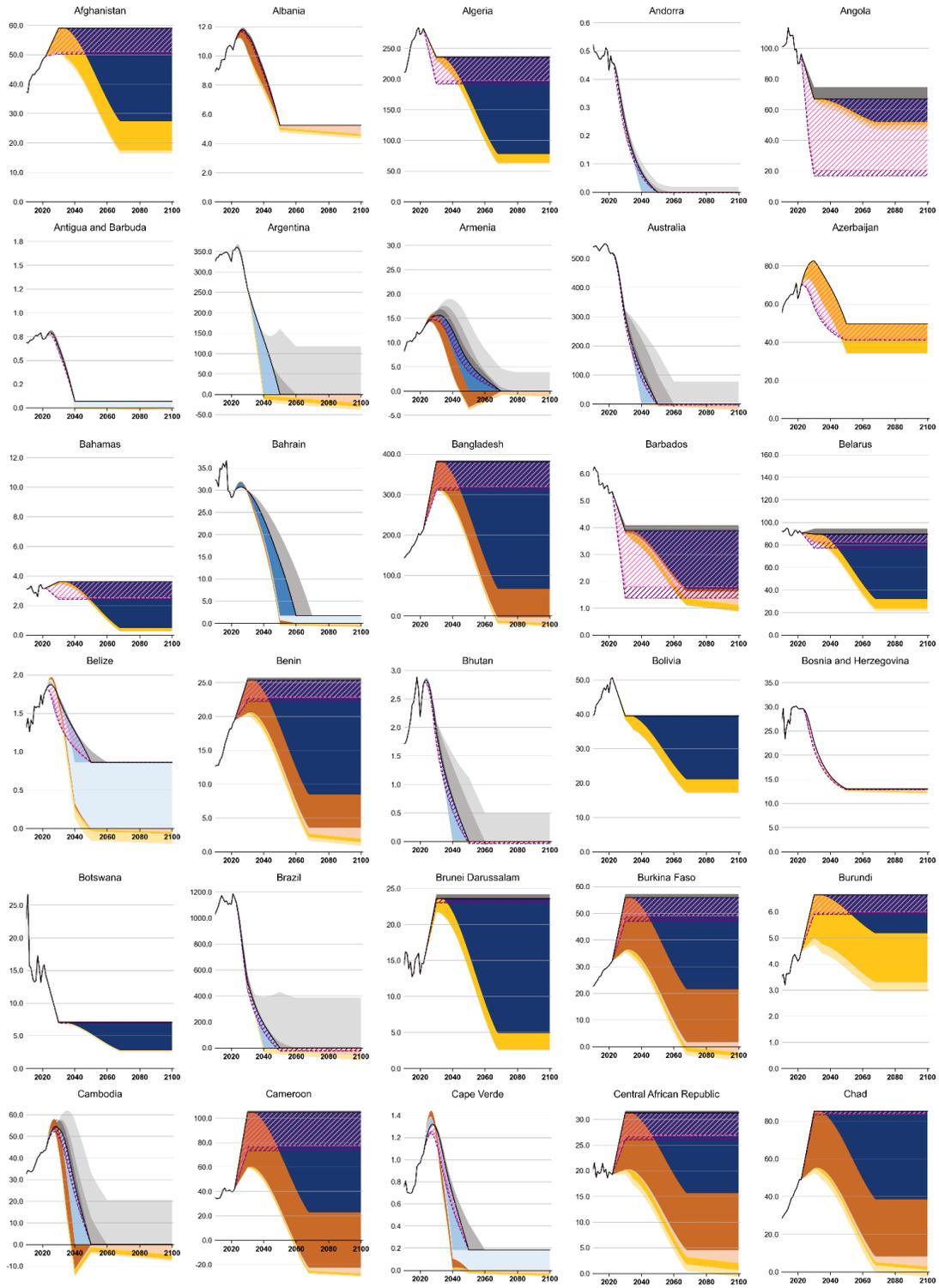

**Fig. S4b. Country-level emissions responses to enhanced or delayed national climate pledges.** Same as Fig. S4 but for other countries and emissions are in Mt/yr.

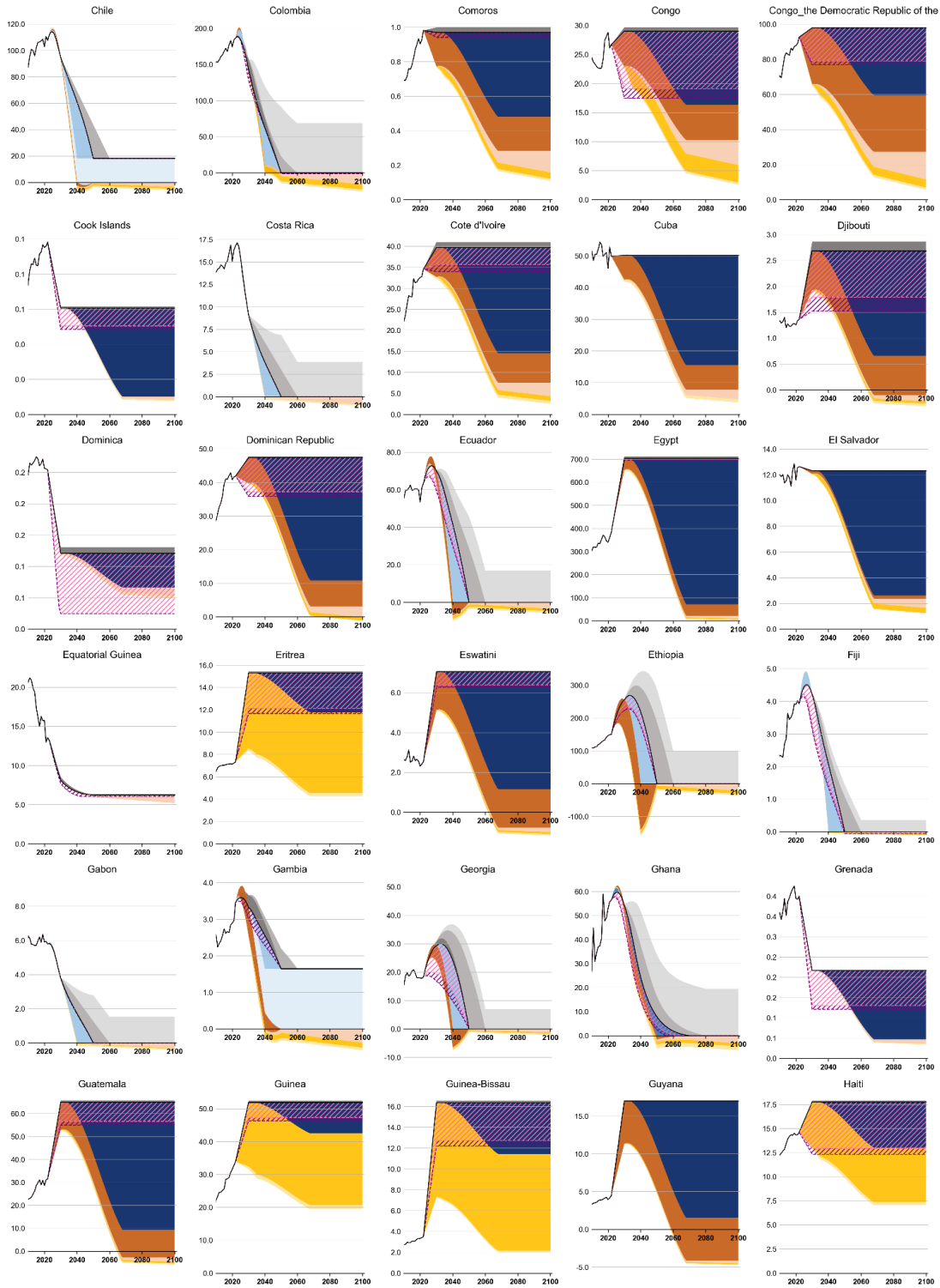

**Fig. S4c. Country-level emissions responses to enhanced or delayed national climate pledges.** Same as Fig. S4 but for other countries and emissions are in Mt/yr.

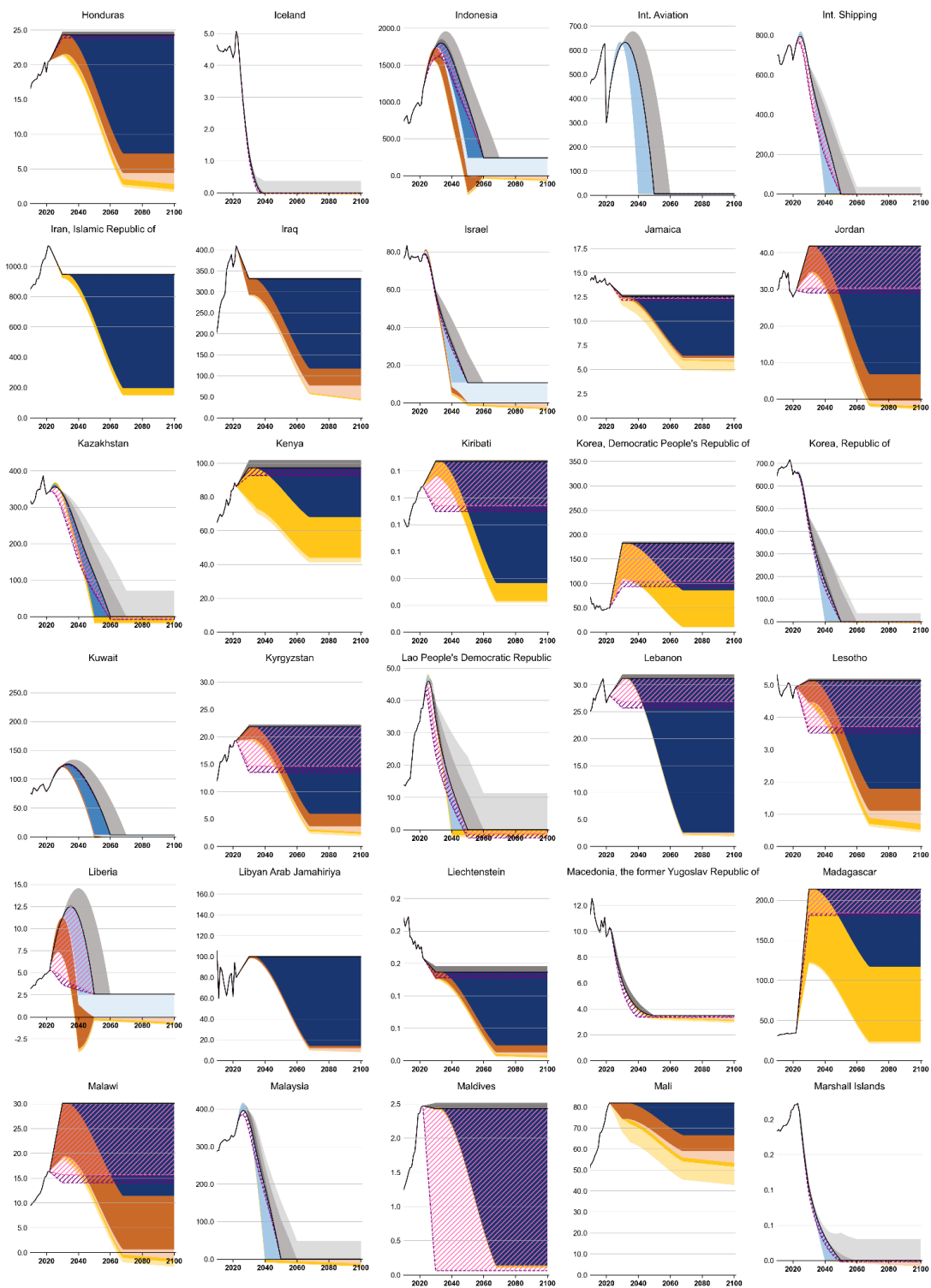

**Fig. S4d. Country-level emissions responses to enhanced or delayed national climate pledges.** Same as Fig. S4 but for other countries and emissions are in Mt/yr.

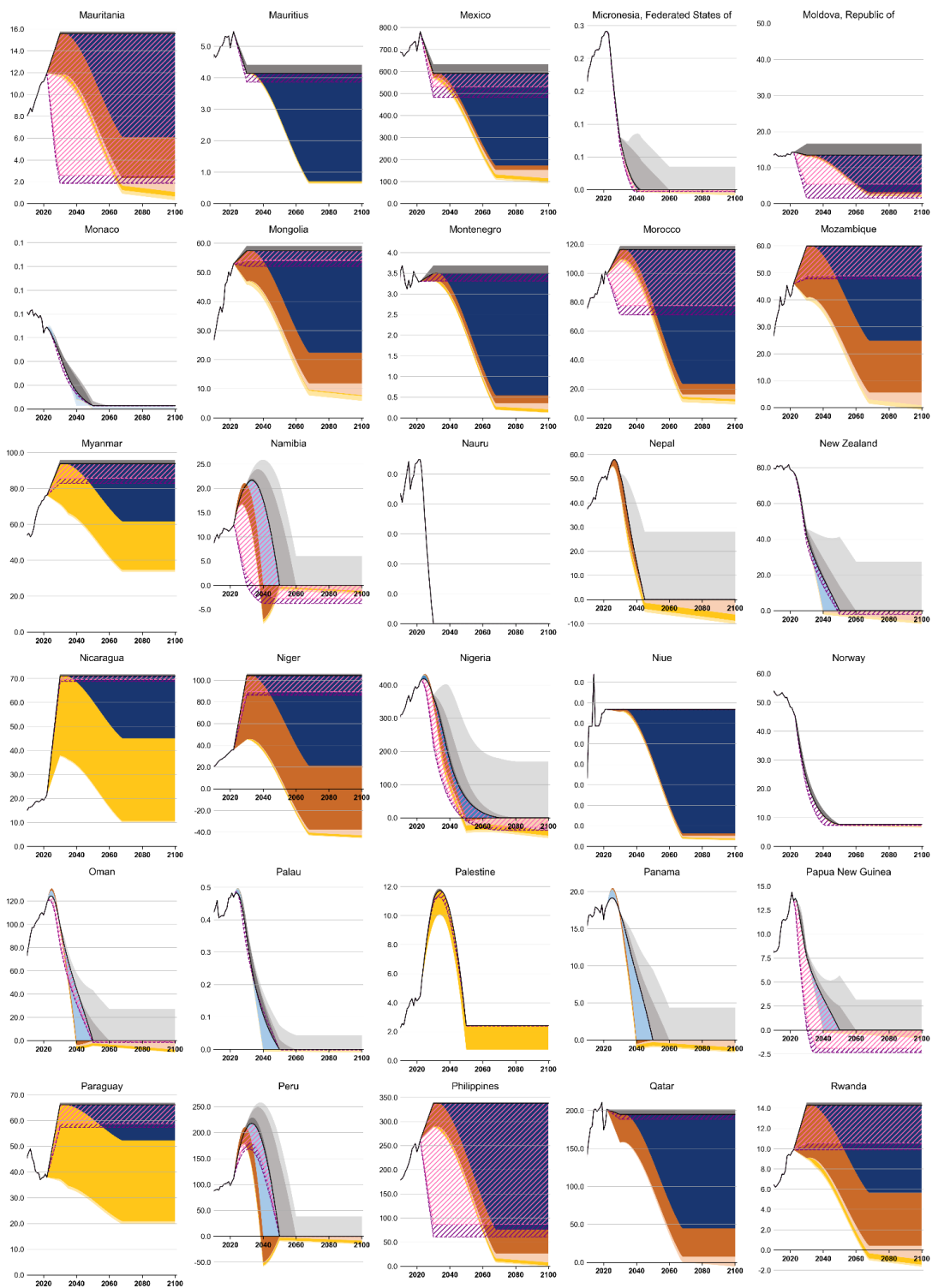

**Fig. S4e. Country-level emissions responses to enhanced or delayed national climate pledges.** Same as Fig. S4 but for other countries and emissions are in Mt/yr.

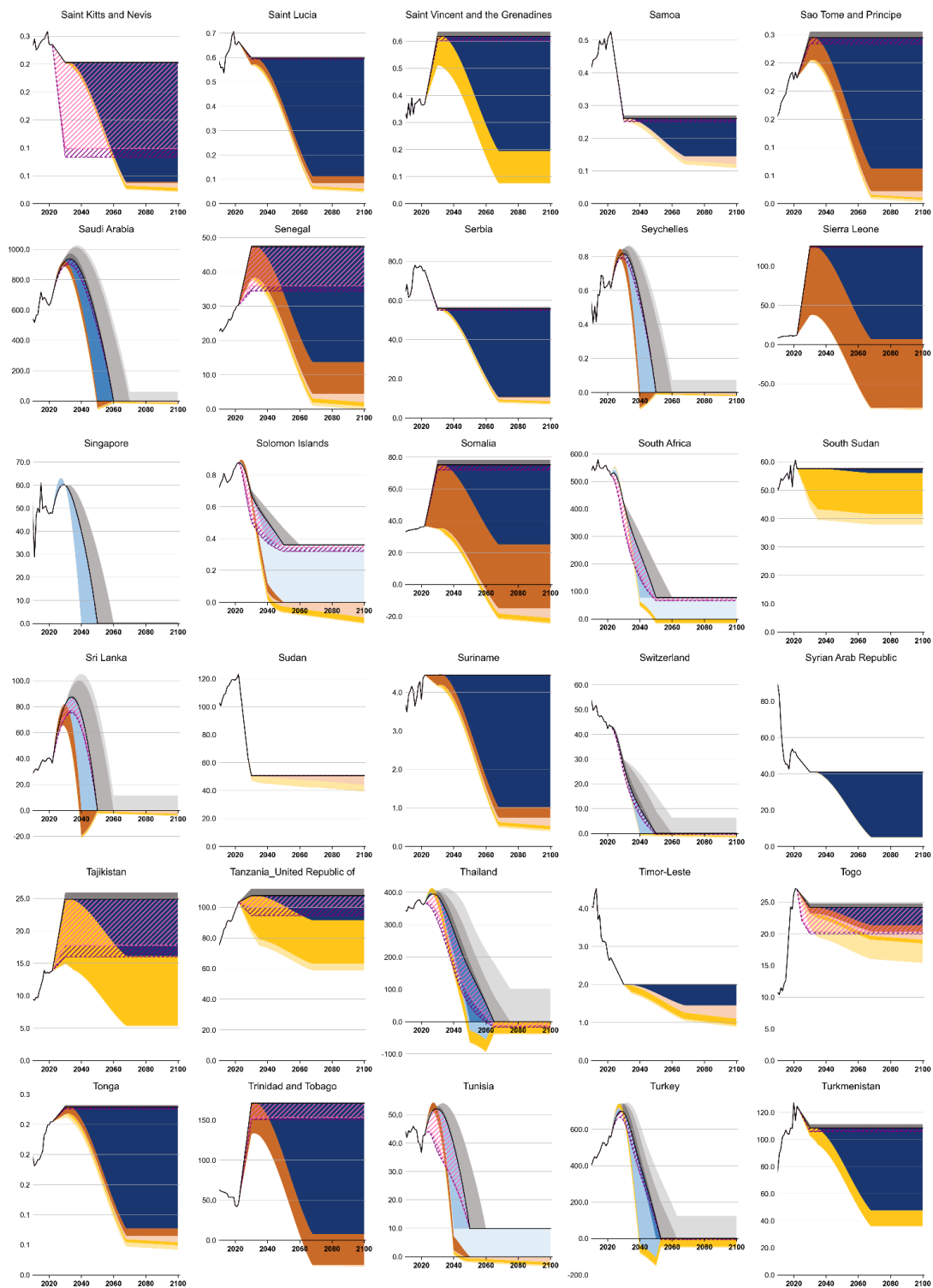

**Fig. S4f. Country-level emissions responses to enhanced or delayed national climate pledges.** Same as Fig. S4 but for other countries and emissions are in Mt/yr.

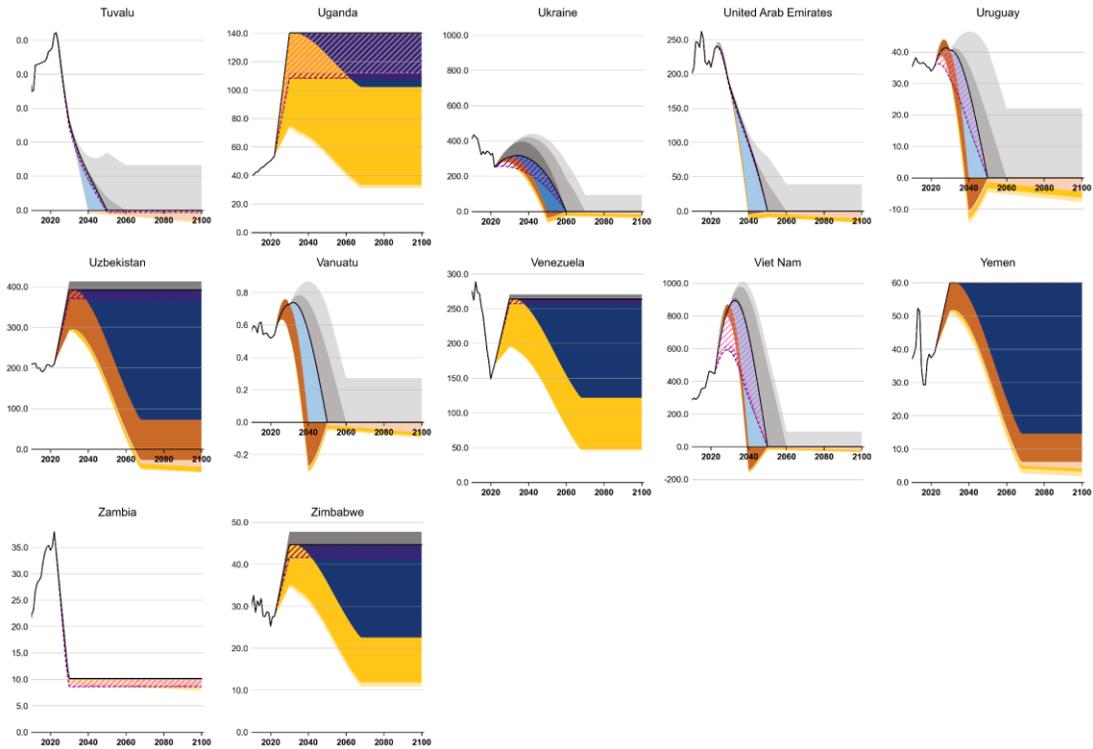

**Fig. S4g. Country-level emissions responses to enhanced or delayed national climate pledges.** Same as Fig. S4 but for other countries and emissions are in Mt/yr.

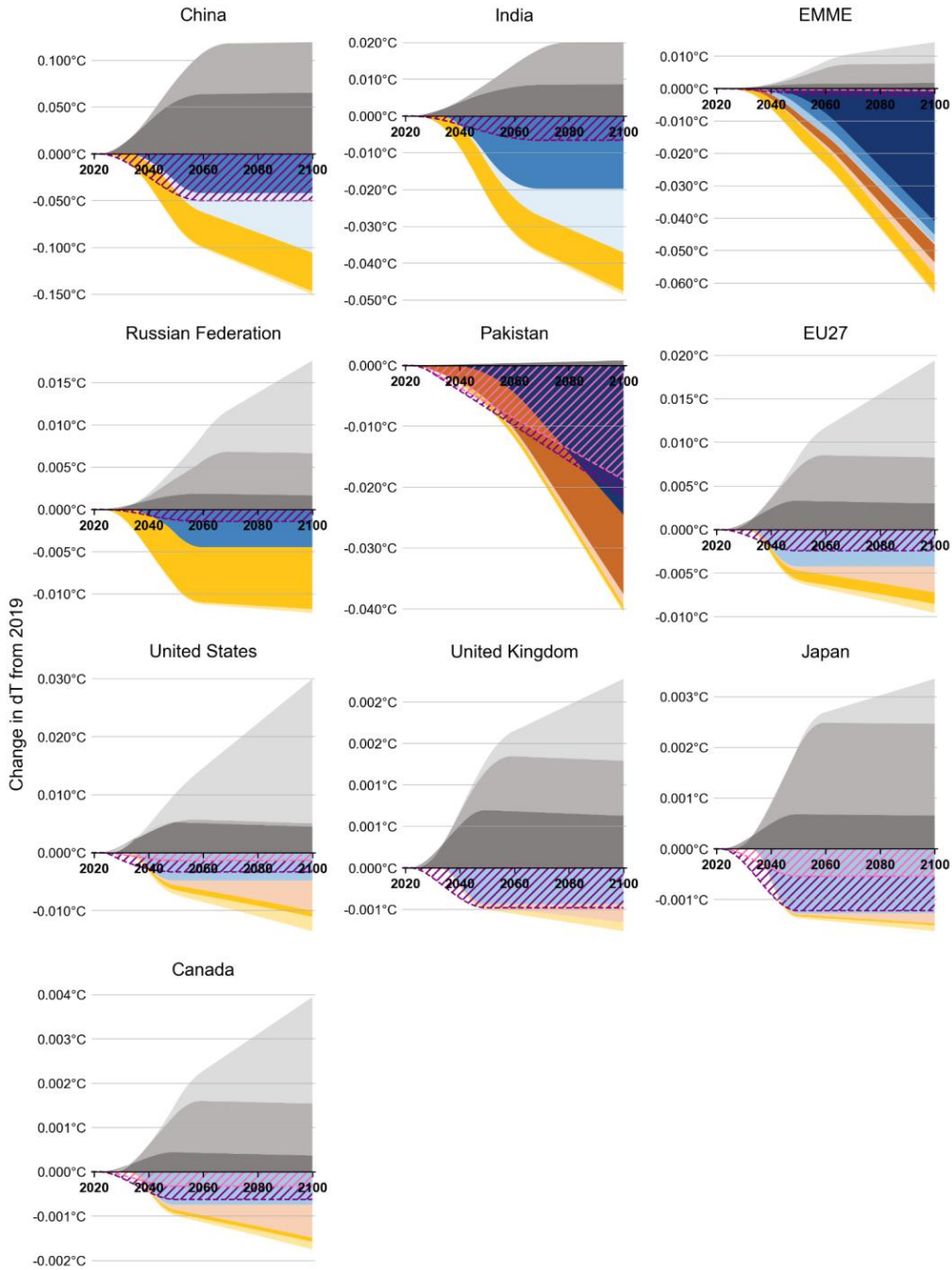

**Fig. S5a. Temperature responses to enhanced or delayed national climate pledges for major emitters.** The figure shows the change in temperatures from 2019 (excluding LULUCF) for each enhanced/delayed pledge compared to temperatures from 2019 for the base case (current pledges with unconditional target). Only for NDC02 (purple shade), the change is with respect to a different base (current pledges with conditional near-term target). Countries are the major emitters from Fig. 1. Responses from LUF pledges are not shown as they are estimated only at the global level. CON and NDC01 pledges (hatched) are shown in a separate layer from other pledges (shaded) due to their potential inter-dependency.

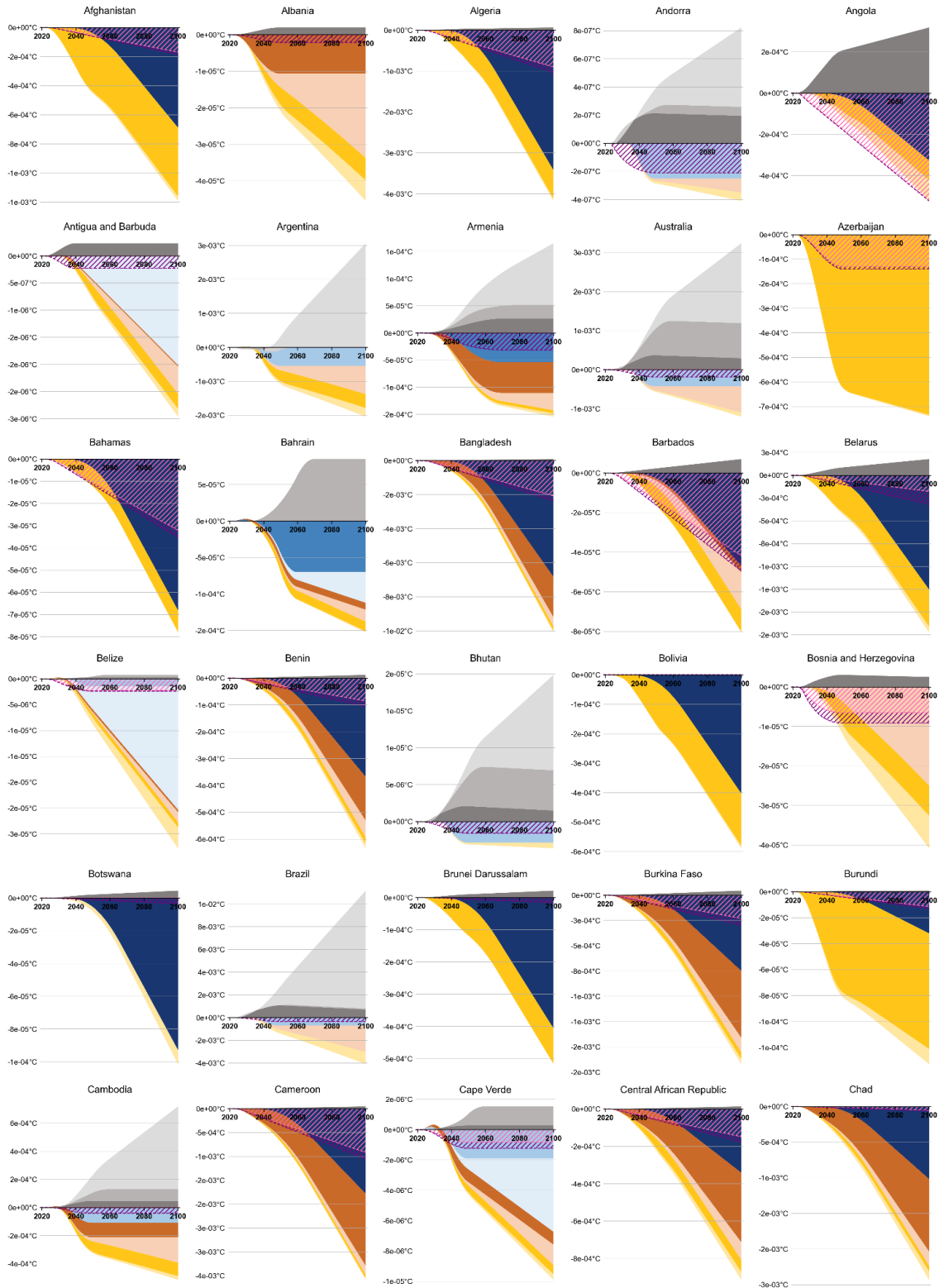

**Fig. S5b. country-level temperature responses to enhanced or delayed national climate pledges.** Same as Fig. S5 but for other countries in alphabetical order.

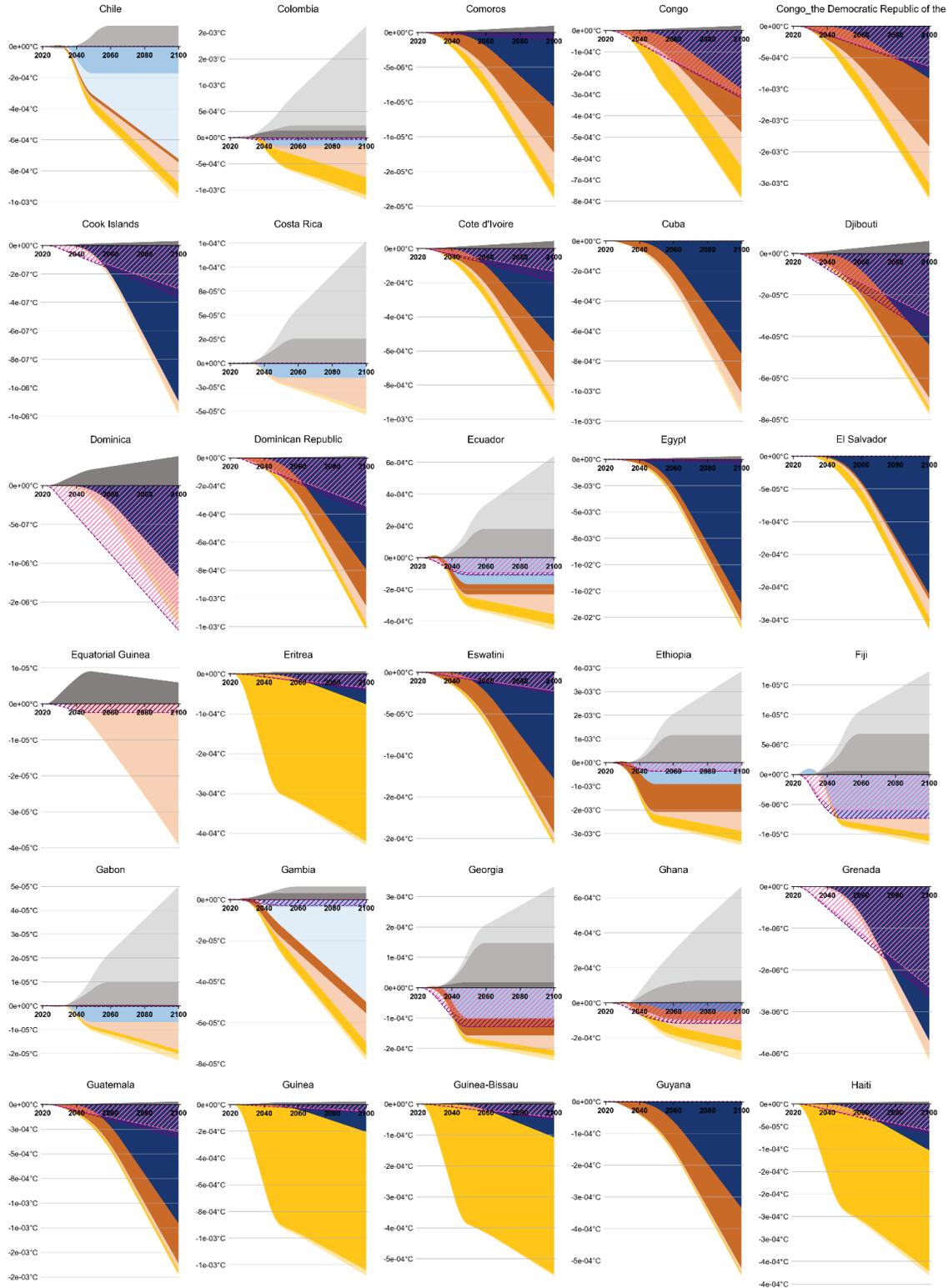

**Fig. S5c. country-level temperature responses to enhanced or delayed national climate pledges.** Same as Fig. S5 but for other countries in alphabetical order.

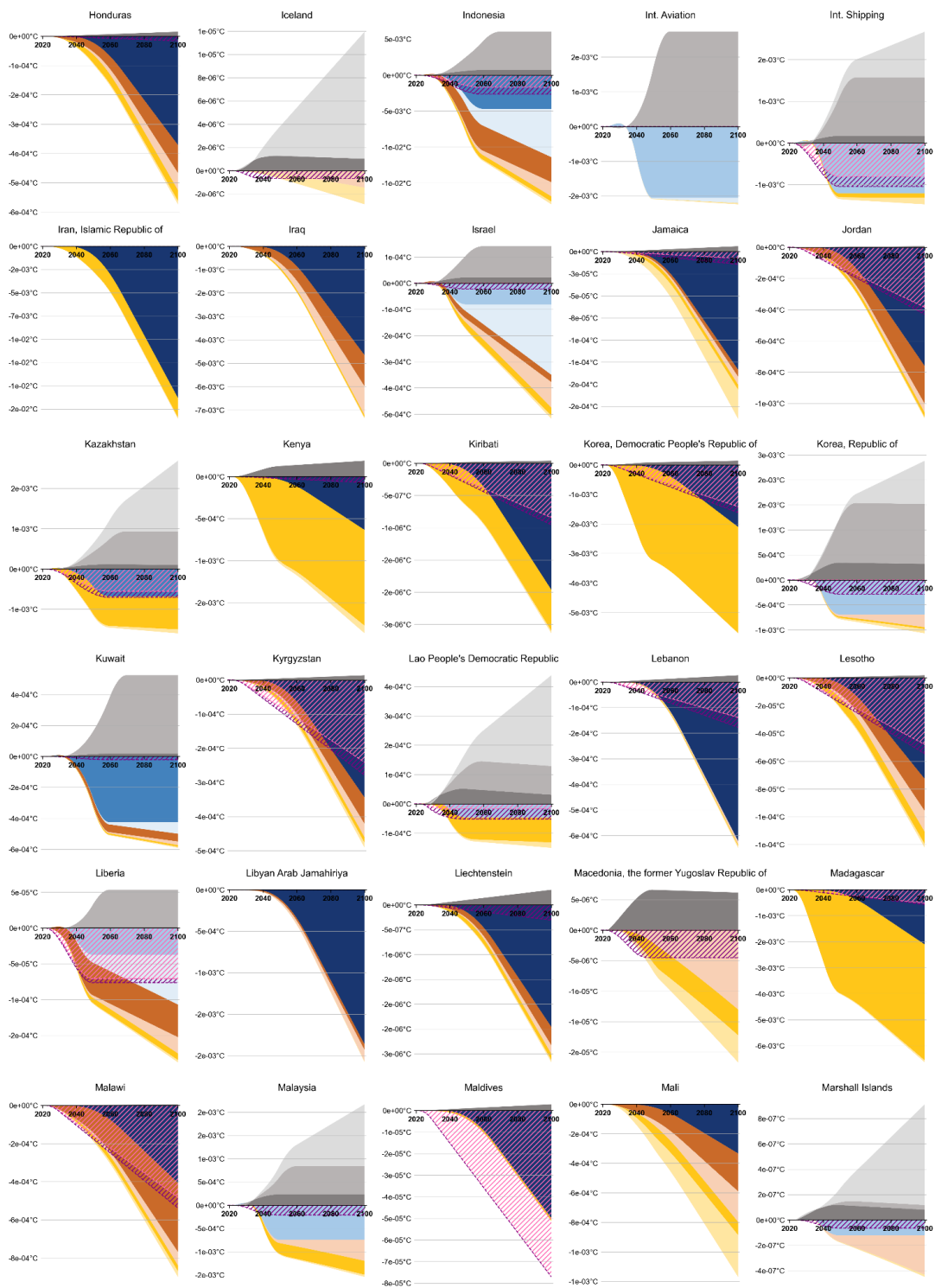

**Fig. S5d. country-level temperature responses to enhanced or delayed national climate pledges.** Same as Fig. S5 but for other countries in alphabetical order.

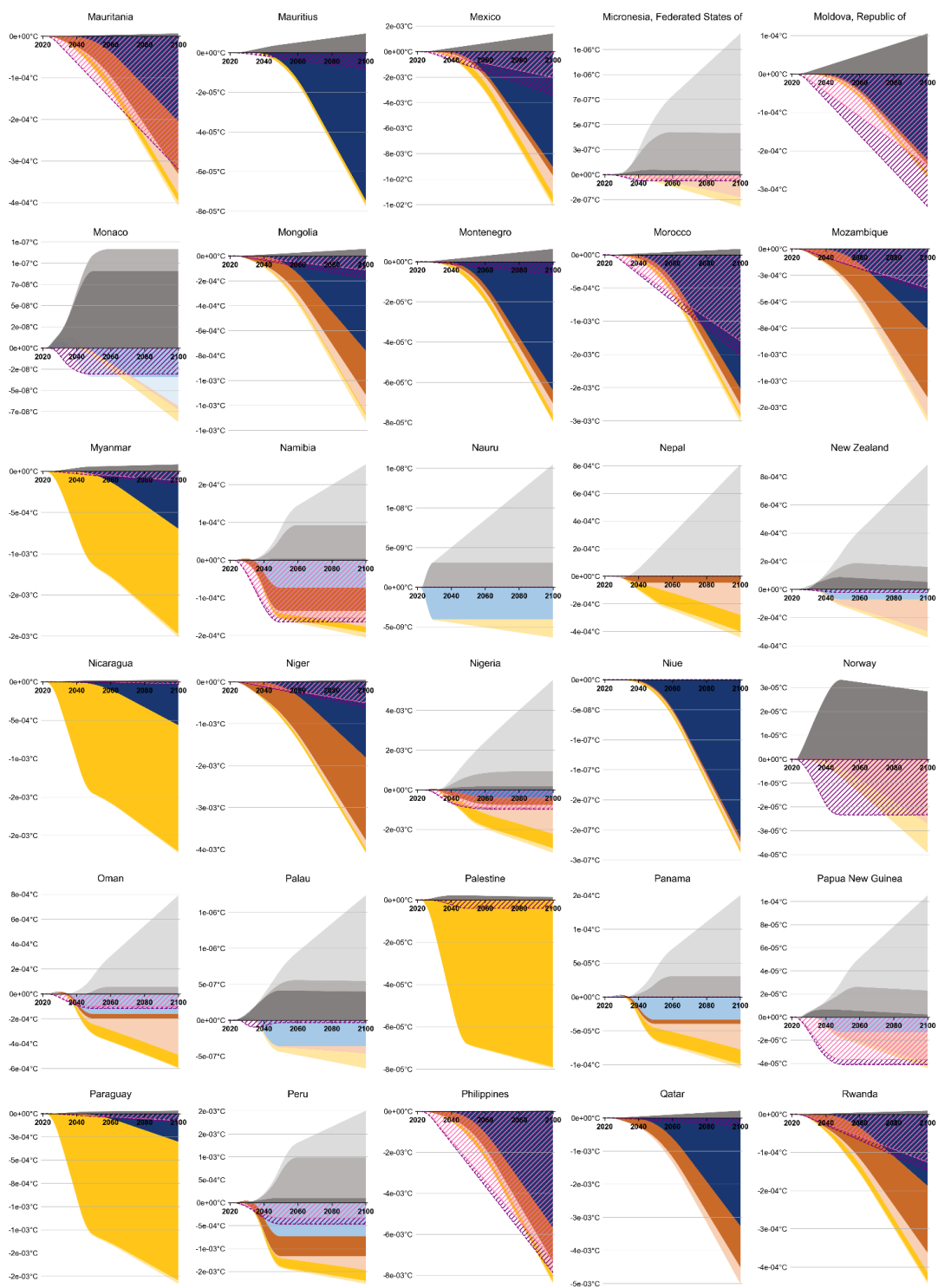

**Fig. S5e. country-level temperature responses to enhanced or delayed national climate pledges.** Same as Fig. S5 but for other countries in alphabetical order.

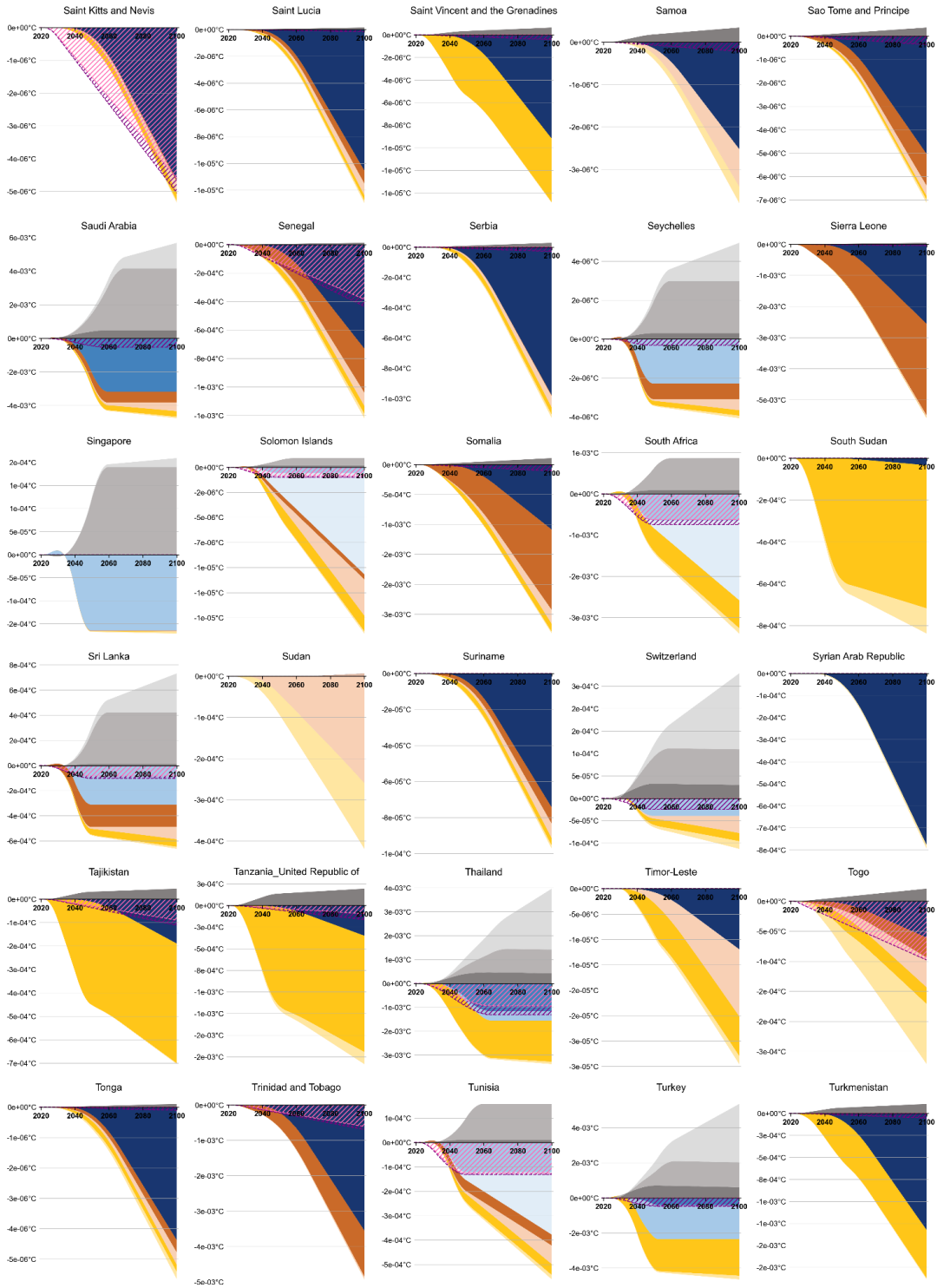

**Fig. S5f. country-level temperature responses to enhanced or delayed national climate pledges.** Same as Fig. S5 but for other countries in alphabetical order.

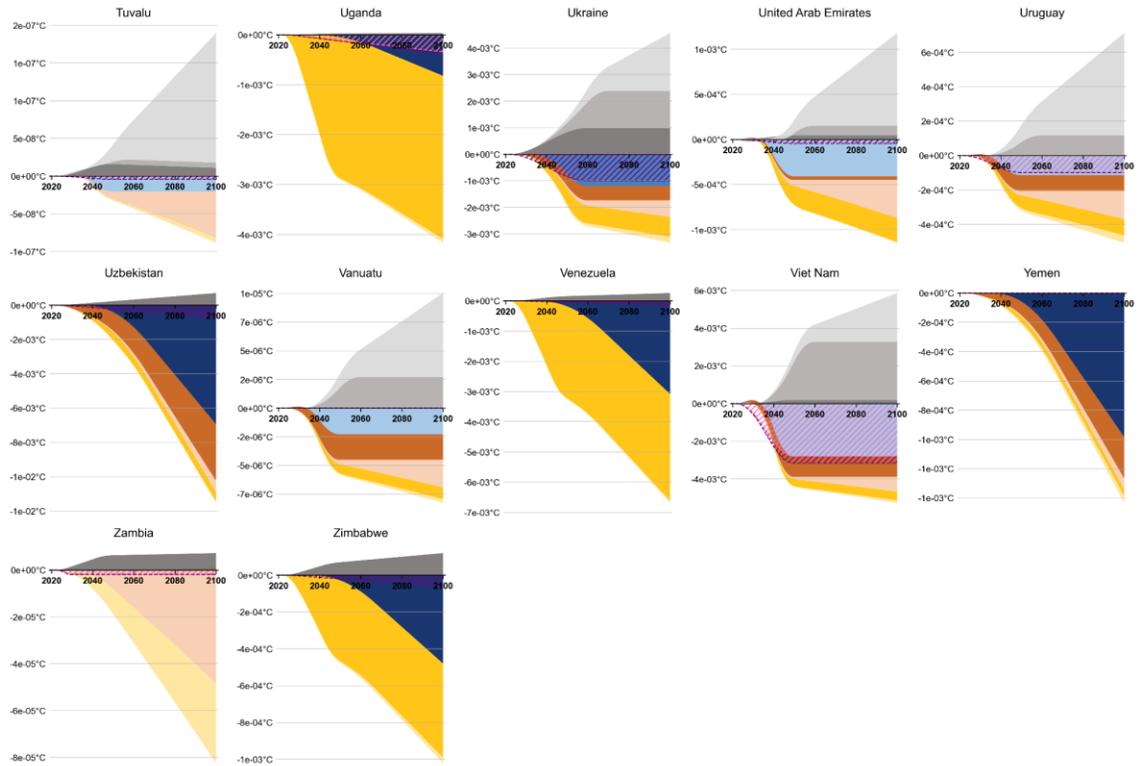

**Fig. S5g. country-level temperature responses to enhanced or delayed national climate pledges.** Same as Fig. S5 but for other countries in alphabetical order.

**Table S1. Summary of experimental setups for LUF pledges.**

| Scenario | For deforestation | For degradation | Reduction level |
|---|---|---|---|
| LUF01 | End Gross Forest loss - BAU_deforestation | 100% control - BAU_degradation | Maximum |
| LUF02 | End Net Forest loss - BAU_deforestation | 50% control - BAU_degradation | Mid-way |
| LUF03 | End Tree Cover loss - BAU_deforestation | 10% control - BAU_degradation | Minimum |

**Table S2. Summary of emission metrics used in the IRF+metric approach.** Note that the values of GWPs and GTPs are based on IPCC AR5 following the Paris Agreement rulebook (*29*). The GWP100 values used in the GWP* formulation also uses AR5 values as described in refs. (*23*, *95*).

| Case | Description |
|---|---|
| m1 : gas by gas | Emissions of $CH_4$ and $N_2O$ are directly used to calculate their respective concentrations without being converted to $CO_2eq$ emissions (i.e., no metric used). |
| m2: GTP100 | Emissions for $CH_4$ and $N_2O$ are converted to $CO_2eq$ emissions using GTP100 (4 and 234, respectively) |
| m3: GWP100 | Emissions for $CH_4$ and $N_2O$ are converted to $CO_2eq$ emissions using GWP100 (28 and 265, respectively) |
| m4: GTP20 | Emissions for $CH_4$ and $N_2O$ are converted to $CO_2eq$ emissions using GTP20 (67 and 277, respectively) |
| m5: GWP20 | Emissions for $CH_4$ and $N_2O$ are converted to $CO_2eq$ emissions using GWP20 (84 and 264, respectively) |
| m6: GWP* | Emissions for $CH_4$ are converted to $CO_2eq$ using GWP-star (GWP*) approach for ensuring best temperature equivalency; emissions for $N_2O$ are converted to $CO_2eq$ using GWP100. Note that we use the notation of $CO_2eq$ for simplicity, although a notation of $CO_2$-warming-equivalent has been proposed for GWP*. |